\documentclass[letterpaper,twocolumn,10pt]{article}
\usepackage{templates/usenix2019,epsfig,endnotes}
\usepackage{fullpage}

\usepackage{hyperref}
\usepackage{url}        
\usepackage{float}
\usepackage{graphicx}
\usepackage{amsmath}
\usepackage{color}
\usepackage{times}
\usepackage{enumitem}
\usepackage[font={scriptsize,it}]{subfig}
\usepackage{verbatim}
\usepackage{xspace}
\usepackage{xfrac}
\usepackage{algorithm, algpseudocode}
\usepackage[normalem]{ulem}
\usepackage{amsfonts}
\usepackage{lastpage}
\usepackage{boxedminipage2e}

\usepackage[export]{adjustbox}

\usepackage[small, compact]{titlesec}
\titlespacing*{\section}{1pt}{3.5pt}{2pt}
\titlespacing*{\subsection}{1pt}{3pt}{1.5pt}
\titlespacing*{\subsubsection}{1pt}{3pt}{1.5pt}

\usepackage[utf8]{inputenc}
\usepackage{cleveref}
\crefname{section}{§}{§§}
\Crefname{section}{§}{§§}
\crefformat{section}{§#2#1#3}
\usepackage{listings}
\usepackage{microtype} 
\usepackage{balance}   
\usepackage{datetime}
\usepackage{morefloats}

\usepackage[font={footnotesize, it}]{caption}
\usepackage{multirow}

\usepackage{tikz}

\DeclareMathVersion{mathchartertext}

\newcommand{\ihat}{\hat{i}}
\newcommand{\secref}[1]{{\S\ref{#1}}}

\renewcommand{\algorithmicrequire}{\textit{Input:}}
\renewcommand{\algorithmicensure}{\textit{Output:}}
\newcommand{\myvec}[1]{\protect\overrightarrow{#1}}

\newcommand{\summation}[2]{\ensuremath{\displaystyle\sum\limits_{#1}^{#2}}}
\include{envdefs}

\definecolor{nodecolor}{RGB}{255,115,115}

\newcommand*\circled[1]{\tikz[baseline=(char.base)]{
\node[shape=circle,draw=nodecolor,inner sep=.3pt, fill=nodecolor] (char) {\color{white}\scriptsize{#1}};}}

\newcommand*\circledd[1]{\tikz[baseline=(char.base)]{
\node[shape=circle,fill=gray,inner sep=.3pt] (char) {\color{white}\scriptsize{#1}};}}

\def\wrt{{w.r.t.}}

\def\vs{vs.}

\newcommand{\cut}[1]{}

\newcommand{\name}{{\textsc{\small Whiz}}}

\newcommand{\Granule}{Capsule}
\newcommand{\Granules}{Capsules}
\newcommand{\granule}{capsule}
\newcommand{\granules}{capsules}

\begin{document}

\title{\bf \textsc{Whiz}: A Fast and Flexible Data Analytics System}
\linespread{1}
\author{
{\rm Robert Grandl, Arjun Singhvi, Raajay Viswanathan, Aditya Akella}\\
University of Wisconsin - Madison\\
} 
\date{}
\maketitle
\thispagestyle{empty}

\noindent
{\bf{\em Abstract---}} Today's data analytics frameworks are
compute-centric, with analytics execution almost entirely dependent on
the pre-determined physical structure of the high-level
computation. Relegating intermediate data to a second class entity in this manner
hurts flexibility, performance, and efficiency. We present \name, a
new analytics framework that cleanly separates computation from
intermediate data. It enables runtime visibility into data via
programmable monitoring, and data-driven computation (where
intermediate data values drive when/what computation runs) via an
event abstraction. Experiments with a \name{} prototype on a large
cluster using batch, streaming, and graph analytics workloads show
that its performance is 1.3-2$\times$ better than state-of-the-art.

\section{Introduction}
\label{sec:introduction}

Many important applications in diverse settings rely on analyzing
large datasets, including relational tables, event streams, and
graph-structured data. To analyze such data, several systems have been
introduced~\cite{hive-icde,sparksql,storm-paper,pregel,graphx,mllib,presto,impala,samza,naiad,ciel}. These
enable data parallel computation, where a job's analysis logic is run
in parallel on data shards spread across multiple machines in large
clusters.

Almost all these systems, be they for
batch~\cite{mapreduce,hive-icde,sparksql},
stream~\cite{storm-paper,samza} or graph
processing~\cite{pregel,graphx,mllib}, have their intellectual roots
in MapReduce~\cite{mapreduce}, a time-tested data parallel execution
framework\footnote{We are {\em not} referring to the MapReduce {\em
    programming model} here.}. While they have many differences, the
systems share a key attribute with MapReduce, in that they are {\em
  compute-centric} (\secref{sec:osdi-motivation}). Their focus, like
MapReduce, is on splitting a job's computational logic, and
distributing it across {\em tasks} to be run in parallel. Like
MapReduce, all aspects of the subsequent execution of the job are
rooted in the job's computational logic, and its task-level
distribution, i.e., the job computation structure. These include the
fact that the compute logic running inside tasks is static and/or
predetermined; intermediate data is partitioned and routed to where it
is consumed based on the compute structure; and dependent tasks are
launched when a fraction of upstream tasks they depend on
finish. These attributes of job execution are not related to, or
driven by, the properties of intermediate data, i.e., how much and
what data is generated. Thus, intermediate data is a {\em second-class
  citizen}.

Compute-centricity was a natural choice for MapReduce. Knowing job
structure beforehand simplifies carving computational units to
execute tasks.  Compute centricity provided clean mechanisms to
recover from failures -- only tasks on a failed machine needed to be
re-executed. Job schedulers became simple because of having to deal
with static inputs, i.e., fixed tasks/dependency structures.
While originally designed for batch analytics, frameworks for streaming and graph analytics have shown that compute-centricity can be applied broadly.

{\em However, given the benefit of hindsight, is compute-centricity
  the right choice?} Our experience with building cluster schedulers,
query optimizers, and execution engines, has shown that atleast four
impediments arise from compute-centricity
(\secref{sec:osdi-motivation}): (1) Intermediate data being an opaque
entity means there is no way to adapt job execution based on runtime
data properties. (2) Static parallelism and intermediate data
partitioning inherent to compute-centric frameworks constrain
adaptation to data skew and resource flux. (3) Execution schedules
being tied to compute structure can lead to resource waste while tasks
wait for input to become available. (4) Compute-based organization of
intermediate data can result in storage hotspots and poor cross-job
I/O isolation, and it curtails data locality.
Thus, compute-centricity begets inflexibility, poor performance, and
inefficiency, which hurts production applications and cluster
deployments.

Instead of adopting point fixes to today's systems and work within
compute-centricity limitations, this paper seeks a ground-up redesign.
We observe that the above limitations arise from (1) tight {\em
  coupling} between intermediate data and compute, and (2) {\em
  intermediate data agnosticity}. Our framework, {\name}, cleanly
separates computation from all intermediate data. Intermediate data is
written to/read from a separate distributed key-value datastore. The
store offers programmable visibility -- applications can provide
custom routines for monitoring runtime data properties.  An event
abstraction allows the store to signal to an execution layer when an
application's runtime data values satisfy certain
properties. Decoupling, monitoring, and events enable {\em data-driven
  computation}: based on data properties, \name{} decides {\em what
  logic} to launch in order to further process data generated, {\em
  how many parallel tasks} to launch, {\em when/where to launch} them,
and {\em what resources} to allocate to tasks. We show that such
data-driven computation helps improve performance, efficiency,
isolation, and flexibility, across a range of different application domains.

We make the following contributions in designing \name: (1) We present
scalable APIs for programmable intermediate data monitoring, and for
applications leveraging events for data-driven actions. Our APIs
balance expressiveness against overhead. (2) We show how to organize
intermediate data from multiple jobs in the datastore so as to achieve
data locality, fault tolerance, and cross-job isolation. Since
obtaining an optimal data organization is NP-Hard, we develop novel
heuristics that carefully trade-off among these objectives. (3) We
develop novel iterative heuristics for the execution layer for
data-driven task parallelism and placement. This minimizes runtime
skew in data processed and lowers data shuffle cost. (4) We launch
each task in a container whose size is late-bound to the actual data
allocated to the task, ensuring low data processing skew and optimal efficiency
under resource dynamics.

We have built a \name{} prototype by modifying
Tez~\cite{ttez} and YARN~\cite{yarn} ($15$K LOC).  We currently
support batch, graph, and stream processing.  We deploy and experiment
with our prototype on a 50 machine cluster in
CloudLab~\cite{cloudlab}. We compare against several state-of-the-art
compute centric (CC) approaches. \name{} improves median ($95\%$-ile)
job completion time (JCT) by $1.3-1.6\times$ ($1.5-2.2\times$) across
batch, streaming, and graph analytics.  \name{} reduces idling by
launching (the right number of appropriately-sized) tasks only when
custom predicates on input data are met, and avoids expensive data
shuffles even for consumer tasks. Under high cluster load, \name{}
offers $1.8\times$ better JCT than CC due to better cross-job data
management and isolation.
\name{}'s data-driven actions enable computation to start much sooner
than CC, leading to $1.3-1.6\times$ better stream and graph analytics median JCTs.
\section{Compute-Centric Vs. Data-Driven}
\label{sec:osdi-motivation}

We begin with a brief overview of today's batch, stream and graph
processing systems (\secref{sec:background}). We then discuss the
three key design principles of \name{}
(\secref{sec:principles}). Finally, we list the performance issues
arising from  {\em compute-centricity} and show how the {\em
  data-driven} design adopted by \name{} overcomes them
(\secref{sec:issues-existing}).

\subsection{Today: Compute-Centric Frameworks}
\label{sec:background}

Production batch~\cite{spark,mapreduce,tez},
stream~\cite{samza,spark,storm} or graph~\cite{powergraph,pregel,graphx}
analytics frameworks support the execution of multiple interdependent {\em
stages} of computation. Each stage is executed simultaneously within different
{\em tasks}, each processing different data shards, or {\em partitions}, to
generate input for a later stage.

\begin{figure}
  \centering
  \subfloat[][A batch analytics job. Intermediate data is partitioned into two key ranges, one per reduce task, and stored in local files at map tasks.]{%
    \label{fig:sample-batch-analytics}%
    \includegraphics[scale=0.5]{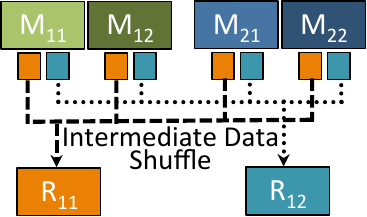}%
  }%
  \hspace{0.2cm}%
  \subfloat[][Data flow in a streaming job. Tasks in all stages are always
  running. Output of a stage is immediately passed to a task in 
  downstream stage. However, CPU is idle until task in Stage 2 receives 100 records
  after which computation is triggered.]{%
    \label{fig:sample-streaming-job}%
    \includegraphics[scale=0.56]{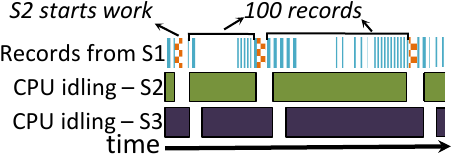}%
  }
  \vspace{-0.1in}
  \caption{Simplified examples of existing analytics systems}
\label{fig:samples}  \vspace{-0.2in}
\end{figure}

Fig.~\ref{fig:sample-batch-analytics} is an example of a simple {\em
  batch analytics} job. Here, two tables need to be filtered based on
provided predicates and joined to produce a new table. There are 3
stages: two maps for filtering and one reduce to perform the
join. Execution proceeds as follows: (1) Map
tasks from both the stages execute first with each task processing a
partition of the corresponding input table. (2) Map intermediate
results are written to local disk by each task, split into files, one
per consumer reduce task. (3) Reduce tasks are launched when the map
stages are nearing completion; each reduce task {\em shuffles}
relevant intermediate data from all map tasks' locations, and
generates output.

A {\em stream analytics} job (e.g.,
Fig.~\ref{fig:sample-streaming-job}) has a similar
model~\cite{storm,heron,spark-streaming}; the main difference
is that tasks in all stages are always running. A {\em graph
  analytics job}, in a framework that relies on the popular message
passing abstraction~\cite{pregel}, has a similar but simplified model:
the different stages are iterations in a graph algorithm, and thus
all stages execute the same processing logic (with the input being the
output of the previous iteration). 

As the above shows, today’s frameworks are designed primarily with the goal of
splitting up and distributing computation across multiple machines,
making them {\em compute centric} (\secref{sec:introduction}). 
  Intermediate data is a second class citizen, strewn across many
  files at the locations where producer tasks are run, and routed
  along predetermined edges between producer and consumer tasks.

\subsection{\name{}: A Data-Driven Framework}
\label{sec:principles}

\name{} makes intermediate data a first class citizen
(Fig.~\ref{fig:f2_flow}). It achieves this by adopting three design
principles:

\noindent \textbf{Decoupling compute and data:} \name{} decouples
compute from intermediate data (\secref{sec:data_org}). Data from all
stages across all jobs is written to/read from a separate key-value (KV)
datastore, and managed by a distinct data management layer called the
data service (DS). An execution service (ES) manages compute tasks.

\noindent \textbf{Programmable data visibility:} The above separation
also enables low-overhead approaches to gain visibility into {\em all}
runtime data (\secref{s:monitor}).  A programmer can gather custom
runtime data properties via a simple API.


\noindent \textbf{Data-driven computation:} Building on data
visibility, \name{} provides an API for applications to act on
\emph{events}. Events form the basis for an {\em intermediate data
  publish-subscribe substrate}. Programmers can define custom
predicates on intermediate data properties for each stage, and events
notify the application when intermediate data satisfies the
predicates. Events help achieve {\em data-driven computation}:
properties of intermediate data drive all aspects of further
computation (\secref{s:da}).

\subsection{Overcoming Issues with Compute-centricity}
\label{sec:issues-existing}

We contrast \name{} with compute centricity along flexibility,
performance, efficiency, placement, and isolation.

\noindent \textbf{Data opacity, and compute rigidity:} In
compute-centric frameworks, there is no visibility into intermediate
data generated by different stages of a job and the tasks’
computational logic are decided apriori. This prevents adapting the
tasks' logic based on their input data.  Consider the job
in Fig.~\ref{fig:sample-batch-analytics}. Existing frameworks
determine the type of join for the {\em entire} reduce stage based on
coarse statistics~\cite{hive}; unless one of the tables is small, a
sort-merge join is employed to avoid out-of-memory (OOM) errors. On
the other hand, having per-key histograms of intermediate data would
enable dynamically determining the type of join to use for {\em
  different} reduce tasks. A task can use hash join if the total size
of {\em its} key range is less than the available memory, and merge
join otherwise.  \name{} offers visibility into all intermediate data
to support computation of such rich statistics which can be used to decide
the logic to apply.

\noindent \textbf{Static Parallelism, Partitioning:} Today, jobs'
per-stage parallelism, inter-task edges and intermediate data
partitioning strategy are decided independent of runtime data and
resource dynamics. In Spark~\cite{spark} the number of tasks in a
stage is determined apriori by the user application or by
SparkSQL~\cite{spark-sql}. A hash partitioner is used to place an
intermediate $(k,v)$ pair into one of $|tasks|$
buckets. Pregel~\cite{pregel} vertex-partitions the input graph;
partitions do not change during the execution of the algorithm.

This limits adaptation to {\em resource flux} and {\em data skew}. A
running stage cannot utilize newly available compute
resources~\cite{qoop,cumulon,cumulond} and dynamically increase its
parallelism. If some key (or some vertex program) in a partition has
an abnormally large number of records (or messages) to process then
the corresponding task is significantly slowed down~\cite{mantri},
affecting both stage and overall job completion times. Data skew is
hard to predict.

With \name{}, because runtime data is managed independently
(\secref{sec:data_org}), compute and parallelism for downstream stages
can be {\em late-bound} (\secref{es:formulation}). Based on the actual
volume of runtime data, and the current resources, we determine how many tasks
to launch and how to provision them. This controls data skew, and
provisions task resources proportional to the data to be processed.

\noindent \textbf{Idling due to compute-driven scheduling:} Modern
schedulers~\cite{yarn,drf} decide when to launch tasks for a stage
based on the static computation structure. When a stage's computation
is commutative+associative, schedulers launch its tasks once 90\% of
all tasks in upstream stages complete~\cite{tez}. But the remaining
10\% producers can take long to complete~\cite{mantri} resulting in
tasks idling.

Idling is worse in streaming, where consumer tasks are continuously
waiting for data from upstream tasks. E.g., consider the streaming job
in Fig.~\ref{fig:sample-streaming-job}. Stage 2 computes and outputs
the median for every 100 records received. Between computation, S2's
tasks stay idle. As a result, the tasks in the downstream S3 stage
{\em also} lay idle. To avoid idling, tasks should be scheduled {\em
  only when, and only as long as, relevant input is available}. Above,
computation should be launched only after $\ge 100$ records have been
generated by an S1 task.

Likewise, in batch analytics, if computation is commutative+associate,
it is beneficial to ``eagerly'' launch tasks to process intermediate
data whenever enough data has been generated to process in one batch,
and exit soon after done. This is challenging to achieve today due to
compute-driven scheduling and lack of data visibility.

Idling is easily avoided with \name{}: programmable monitoring aids
runtime data statistics collection; when statistics indicate that
relevant data has been generated, events are triggered, aiding launch
of relevant tasks.

\noindent \textbf{Placement, and storage isolation:} Because
intermediate data is spread across producer tasks' locations, it is
impossible to place {\em consumer tasks} in a data-local fashion. Such
tasks are placed at random~\cite{mapreduce} and forced to engage in
expensive shuffles that consume a significant portion of job run times
($\sim$30\%~\cite{orchestra}).

Also, when tasks from multiple jobs are collocated, it becomes
difficult to isolate their hard-to-predict runtime intermediate data
I/O. Tasks from jobs generating large intermediate data may occupy
much more local storage and I/O bandwidth than those generating less.

Since the \name{} store manages data from all jobs, it can enforce
policies to organize data to meet {\em per-job} objectives, e.g., data
locality for {\em any} stage (not just input-reading stages), {\em
  and} to meet {\em cluster} objectives, such as I/O hotspot avoidance
and cross-job isolation.

\subsection{Related Work}
\label{sec:related}


\noindent \textbf{Data opacity:}

Almost all database and bigdata SQL
systems~\cite{rope,shark,spark-sql} use statistics computed ahead of
time to optimize execution. Adaptive query optimizers
(QOs)~\cite{aqp-deshpande} use dynamically collected statistics and
re-invoke the QO to re-plan queries top-down. In contrast, \name{}
alters the query plans on-the-fly at the {\em execution layer} based
on {\em run-time} data properties, thereby circumventing additional
expensive calls to the QO.  Optimus~\cite{optimus} allows changing
application logic based on approximate statistics operators that are
deployed alongside tasks. However, the system targets simple
computation logic rewriting, and cannot enable other data-driven
benefits, e.g., adapting parallelism, and rightsizing  tasks.

\noindent \textbf{Skew handling and static parallelism:} Some
parallel
databases~\cite{bucket-spreading,partition-tuning,parallel-joins} and
big data systems~\cite{skewtune} dynamically adapt to data skew for
single large joins. In contrast, \name{} holistically solves data
skew for all joins across multiple jobs and further strives to achieve
data locality.  \cite{skewtune,partition-tuning} deal with skew in
MapReduce by dynamically splitting data for slow tasks into smaller
partitions and processing them in parallel. But, they can cause
additional data movement from already slow machines leading to poor
performance. \cite{hurricane} mitigates skew by cloning slow
tasks and adaptively partitioning work. However, it has no data
visibility and its data organization does not consider data locality
and fault tolerance.

\noindent \textbf{Decoupling:} Naiad~\cite{naiad} and
StreamScope~\cite{streamscope} also decouple intermediate data. They
tag intermediate data with vector clocks which are used to trigger
compute in the correct order. Thus, both support ordering driven
computation, orthogonal to data-driven computation in
\name{}. Also, Naiad assumes entire data fits in memory. StreamScope
is not applicable to batch/graph analytics.

\noindent \textbf{Storage inefficiencies:} For batch
analytics,~\cite{locus, pywren} addresses storage inefficiencies by
pushing intermediate data to the appropriate external data services
(like Amazon S3, Redis) while remaining cost efficient and running on
serverless platforms. Similarly,~\cite{pocket} is an elastic data
store used to store intermediate data of serverless
applications. However, since this data is still opaque, and
compute and storage are managed in isolation, these systems cannot
support data-driven computation or achieve data locality and load
balancing simultaneously.
\section{\name{} Programming Model}
\label{sec:programming-model}

\begin{table*}
  \centering
  \begin{scriptsize}
    \begin{tabular}{@{}p{0.001cm}|p{5.5cm}|p{10.2cm}@{}}
      & \textbf{API} & \textbf{Description} \\ \hline
    1 & createJob(name:Str, type:Type) & Creates a new job which can be of type BATCH, STREAM or GRAPH. \\ \hline
    2 & createStage(j:Job, name:Str, impl:StageImpl,
    trigger:StageDataReadyTriggerImpl) & Adds a stage of
    computation to a Job with a custom implementation. Optionally, users can specify custom predicates that determine when
    downstream stages can consume current stage's data through
    StageDataReadyTriggerImpl. Otherwise default triggers are applied depending on job type.\\ \hline 
    3 & addDependency(j: Job, s1: Stage, s2: Stage)  & Adds a starts before
    relationship between stages s1 and s2. \\ \hline
    4 & addModifyAction(j:Job, s:Stage, impl:ModifyImpl) & Changes stage computation
    logic as specified by ModifyImpl. ModifyImpl is called when a statEvent arrives to rewrite job description.
    \\ \hline
    5 & replaceStage(j:Job, s:Stage, impl:AlterStageImpl) & Replaces stage implementation with an alternative implementation logic
    \\ \hline
    6 & addDataMonitor(j:Job, s:Stage, impl:DataMonitorImpl) & Adds a data
    monitor to compute statistics over data generated by stage {\em s}. DataMonitorImpl can either be a built-in module or a customized one. 
  \end{tabular}
 \vspace{-1em}
  \end{scriptsize}
  \caption{\name{} Programmer APIs. \label{tab:new-apis}}
  \vspace{-0.2in}
\end{table*}

\begin{figure}[t!]
    \centering
    \includegraphics[clip,width=\columnwidth]{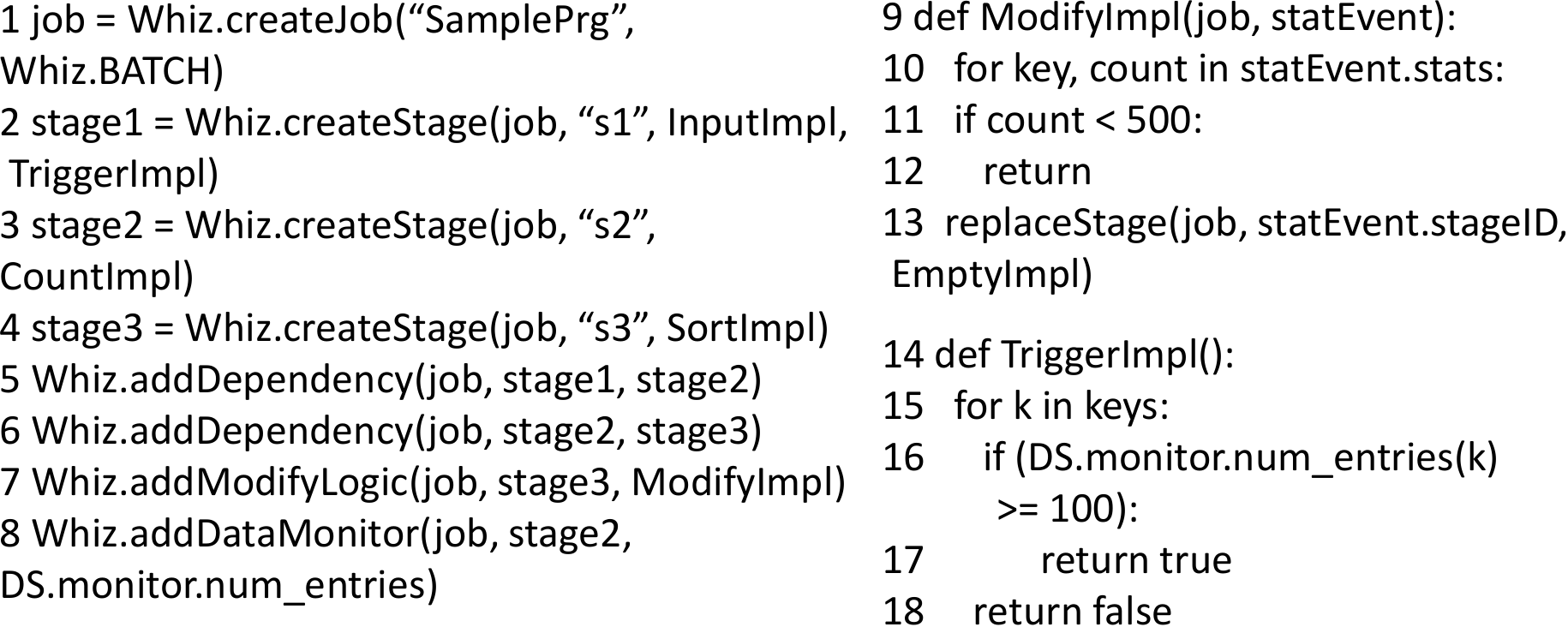}
    \vspace{-1em}
    \caption{An example three-stage job in \name{}}
    \label{fig:new-program}
    \vspace{-0.2in}
\end{figure}


\name{} extends the DAG-based programming model of existing
frameworks~\cite{mapreduce,ttez,storm,pregel} in a few key ways to
support data-driven computation. Contrary to existing frameworks,
\name{} does not require users to provide low-level details such as
stage parallelism and data partitioning strategy.  The minimal
embellishment to familiar DAG-based programming, and freedom from low
level details, ensure that \name{} is simple to program atop.

In \name{}, using the $addDataMonitor$ API (Table.~\ref{tab:new-apis}
row 6), a user can add a built-in or custom module that computes
statistics over data generated by a stage in a data-parallel
program. Using the $createStage$ API (Table.~\ref{tab:new-apis} row
2), a user can provide predicates on the collected data properties
($StageDataReadyTriggerImpl$) to determine when a downstream stage can
consume the current stage's data. Using the $addModifyAction$ API
(Table.~\ref{tab:new-apis} row 4), computation logic can be changed at
runtime based on data properties.

As an example, consider a 3-stage job that processes words and, for
words with $<500$ occurrences, sorts them by frequency. The program
structure is $S_{1} \rightarrow S_{2} \rightarrow S_{3}$, where
$S_{1}$ processes input words, $S_{2}$ computes word occurrences, and
$S_{3}$ sorts the ones with $<500$ occurrences. In \name{} it can be
realized as shown in Fig.~\ref{fig:new-program}. Job composition
details (lines 1-6) are similar to existing frameworks. We specify the
implementation of TriggerImpl for $S_{1}$ and ModifyImpl for $S_{3}$
that help realize data-driven computation. TriggerImpl specifies that
as soon as 100 occurrences of a word are seen, $S_{2}$ can start
running (lines 14-18), facilitating pipelined execution of $S_{1}$ and
$S_{2}$. With the ModifyImpl for $S_{3}$ (lines 9-13), if the data
consists of only words $\ge500$ occurrences, then $S_{3}$ doesn't need
to execute (compute is replaced by null operation). 

In Appendix.~\ref{sec:appendix-programs}, we show several other
example applications written in \name{}. Our prototype supports batch SQL analytics, graph algorithms, and streaming.

\begin{figure}[t!]
    \centering
    \includegraphics[clip,width=\columnwidth]{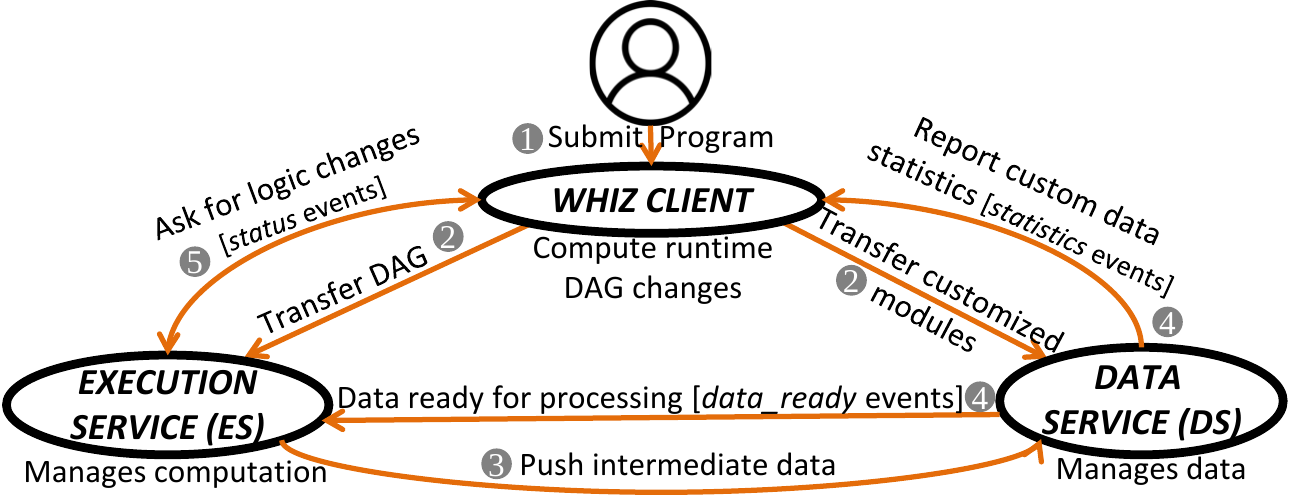}
    \vspace{-1em}
    \caption{\name{} control flow.}
    \label{fig:f2_flow}
    \vspace{-0.2in}
\end{figure}

The overall control flow in \name{} is shown in
Fig.~\ref{fig:f2_flow}. The user program is provided to the \name{}
client. The execution service (ES) runs the root vertex and writes its
output to the datastore (step 3). The Data Service (DS) stores the
received data and notifies ES when data is ready for further
processing via an event (step 4).  DS sends data statistics (e.g.,
per-key counts above) to \name{} client via events as well (step
4). On receiving a data-ready event, the ES (step 5) queries the
client for compute changes. The above process repeats. Interactions between the ES, DS, and the client are hidden from the
user.

\section{Data Store}
\label{sec:data_org}

In \name{}, all jobs' intermediate data is written to/read from a
separate data store, where it is structured as $<$key,value$>$
pairs. In batch/stream analytics, the keys are generated by the stage
computation logic itself; in graph analytics, keys are identifiers of
vertices to which messages (values) are destined for processing in the
next iteration.  We now address how this data is organized in the
store by the data service (DS). The DS does this organization via a cluster-wide master (DS-M).  

An ideal data organization should achieve three cross- and per-job
goals: (1) it should {\em load balance} and spread all jobs' data,
specifically, avoid hotspots and {\em improve cross-job isolation},
and {\em minimize within-job skew} in tasks' data processing. (2) It
should maximize job {\em data locality} by co-locating as much data
having the same key as possible. (3) It should be {\em fault tolerant}
- when a storage node fails, recovery should have minimal impact on
job runtime. Next, we describe our data storage granularity, which forms
the basis for meeting our goals.

\subsection{\Granule: A Unit of Data in \name{}}
\label{s:data_format}


\name{} groups intermediate data based on $<$key$>$s into groups called
\granules{}. A stage's intermediate data is organized into some large
number $N$ \granules{}; crucially $N$ is late-bound as described
below, which helps meet our goals above. Intermediate data key range
is split $N$-ways, and each \granule{} stores all $<$k, v$>$ data from
a given range.  {\name} strives to materialize all \granule{} data on
one machine; rarely, a \granule{} may be spread across a small number
of machines.  This
materialization property of \granules{} forms the basis for consumer
task data locality. In contrast, in today's systems, data from the same key
range may be written at different producers' locations. 


Also, today, key ranges of the intermediate data partitions are
tied to pre-determined task parallelism, whereas \name{} \granule{}
key ranges are {\em unrelated to compute structure} and {\em
  late-bound}. Specifically, we first determine the set of machines on
which \granules{} from a stage are to be stored; machines are chosen
to maximize the ability to simultaneously support isolation, load
balancing, locality, and fault tolerance; the choice of machines then
determines the number $N$ for a stage's \granules{}
(\secref{s:solution}).

Given these machines and $N$, as the stage produces data at run time,
$N$ \granules{} are materialized, and dynamically allocated to
right-sized tasks; this is done in a way that preserves data-local
processing, lowers skew, and optimally uses compute resources
(\secref{sec:runtime_management}).

Thus late-binding realizes an objectives-driven
partitioning and consumption of intermediate data.

\subsection{Fast \Granule{} Allocation}
\label{s:solution}
\label{ss:heuristics}


We consider how to place multiple jobs' \granules{} on machines to
avoid hotspots, ensure data locality and
minimize job runtime impact on data loss.  We formulate a binary
integer linear program (see Fig.~\ref{fig:formulation} in Appendix.
~\ref{s:formulation}) to this end.  Solving this ILP at scale can
take several tens of seconds delaying \granule{} placement.

\name{} instead uses a simpler, practical approach for the \granule{}
placement problem.  First, instead of jointly optimizing global placement
decisions for all the \granules{}, \name{} solves a ``local'' problem of
placing \granules{} for each stage independently (while still considering
inter-stage dependencies); when new stages arrive, or when existing
\granules{} may exceed job quota on a machine, new locations for some of these
\granules{} are determined.  Second, instead of solving a multi-objective
optimization, \name{} uses a linear-time rule-based heuristic to place
\granules{}; the heuristic prioritizes load and locality (in that order) in
case machines satisfying all objectives cannot be found. Isolation (quota) is
always enforced.

\begin{table}[!t]
	\centering
	\begin{scriptsize}
		\begin{tabular}{c|c}
 		\circled{h1} & \multicolumn{1}{|l}{\begin{tabular}[l]{@{}l@{}}\textit{// $Q_{j}$: $\max$ storage quota per job $j$ and machine $m$.} \\ Based on fairness considerations across all runnable jobs $J$.\end{tabular}} \\
		\cline{1-2}
		 & \multicolumn{1}{|l}{\textit{\begin{tabular}[l]{@{}l@{}}// $M_{v}$: number machines (out of $M$) to organize data that will be \\ //generated by $v$ of $j$. \\ \end{tabular}}} \\
		\circled{h2} & \multicolumn{1}{|l}{\begin{tabular}[l]{@{}l@{}} a. Count  number machines $M_{j75}$ where  $j$ is using $<75\%$ of $Q_{j}$; \\ b. $M_{v}$ = $max(2, M_{j75} \times \frac{M_ - M_{j75}}{M})$. \end{tabular}} \\		
		\cline{1-2}
		& \multicolumn{1}{|l}{\textit{\begin{tabular}[l]{@{}l@{}}// Given $M_{v}$, compute list of machines $\myvec{M_{v}}$\end{tabular}}.} \\
		\circled{h3} & \multicolumn{1}{|l}{\begin{tabular}[l]{@{}l@{}} Considers only machines  where $j$ is using $<75\%$ of $Q_{j}$; \\ a. Pick machines that provide LB\footnote{Load balancing}, DL\footnote{Data locality} and maximum possible FT\footnote{Fault tolerance};\\ b. If $|\myvec{M_{v}}|$ $< M_{v}$, relax FT guarantees and pick machines that \\ provide LB and DL; \\ c.  If $|\myvec{M_{v}}|$ is still $< M_{v}$, pick machines that just provide LB. \end{tabular}} \\
		\cline{1-2}
		 & \multicolumn{1}{|l}{\textit{\begin{tabular}[l]{@{}l@{}}// Given $M_{v}$, compute total \granules{} N.\end{tabular}}} \\
		\circled{h4} & \multicolumn{1}{|l}{\begin{tabular}[l]{@{}l@{}}$N$ = $G$ X $M_{v}$, where $G$ = \granules{} per machine \end{tabular}} \\
		\cline{1-2}
		 & \multicolumn{1}{|l}{\textit{\begin{tabular}[l]{@{}l@{}}// Which machines are at risk of violating $Q_{j}$?\end{tabular}}} \\
		\circled{h5} & \multicolumn{1}{|l}{\begin{tabular}[l]{@{}l@{}}$\myvec{M_{j}}$: machines which store data of $j$ and $j$ is using $\ge75\%$ of $Q_{j}$. \end{tabular}} \\
		\cline{1-2}
		 & \multicolumn{1}{|l}{\textit{\begin{tabular}[l]{@{}l@{}}// Which \granules{} are hot on $\myvec{M_{j}}$?\end{tabular}}} \\
		\circled{h6} & \multicolumn{1}{|l}{\begin{tabular}[l]{@{}l@{}}Significantly larger in size or  have a higher increasing rate than others. \end{tabular}} \\
		\end{tabular}
		  \caption{Heuristics employed in data organization}
		  \label{tab:heuristics}
                  \vspace{-0.2in}
	\end{scriptsize}
\end{table}

\label{ss:initial_organization}

\noindent
{\bf \Granule{} location for new stages:} 
When a job $j$ is ready to run, DS-M invokes an admin-provided
heuristic \circled{h1} (Table~\ref{tab:heuristics}) that assigns $j$ a
quota $Q_{j}$ per machine.

When a stage $v$ of job $j$ starts to generate intermediate data, DS-M
invokes \circled{h2} to determine the number of machines $M_{v}$ for
organizing $v$'s data. \circled{h2} picks $M_{v}$ between $2$ and a
fraction of the total machines which are $\le 75\%$ of the quota
$Q_{j}$ for $j$. $M_{v} \ge2$ ensures opportunities for data parallel
processing (\secref{es:formulation}); a bounded $M_{v}$
(Table~\ref{tab:heuristics}) controls the ES task launch overhead
(\secref{es:formulation}). 


Given $M_{v}$, DS-M invokes \circled{h3} to generate a list of machines
$\myvec{M_{v}}$ to materialize data on. It starts by creating three
sub-lists: (1) For load balancing, machines are sorted
lightest-load-first, and only ones which are $\le 75\%$ quota usage
for the corresponding job are considered. (2) For data locality, we
prefer machines which already materialize other \granules{} for this
stage $v$, or \granules{} from other stages whose output will be
consumed by same downstream stage as $v$ (e.g., the two map stages in
Fig.~\ref{fig:sample-batch-analytics}). (3) For fault tolerance, we
pick machines where there are no \granules{} from any of $v$'s $k$
upstream stages in the job, sorted in descending order of
$k$. Thus, for the largest value of $k$, we have all machines
  that do not store data from {\em any} of $v$'s ancestors; for $k=1$
  we have nodes that store data from the immediate parent of $v$. 

We pick machines from the sub-lists to maximally meet our objectives
in 3 steps: (1) Pick least loaded machines that are data local and
offer as {\em high fault tolerance as possible} (machines present in
all three sub-lists). Note that as we go down the fault tolerance list
in search of a total of $M_{v}$ machines, we trade-off fault
tolerance. (2) If despite reaching minimum fault tolerance as
possible, i.e., reaching the bottom of the fault tolerance sub-list --
the number of machines picked falls below $M_{v}$, we completely
trade-off fault tolerance and pick least loaded machines that are data
local (machines present in load balancing and data local
sub-lists). (3) If still the number of machines picked falls below
$M_{v}$, we trade-off data locality and simply pick
least-loaded machines.

Finally, given $\myvec{M_{v}}$, DS-M invokes \circled{h4} and
instantiates a fixed number ($G$) of \granules{} per machine leading
to total \granules{} per-stage ($N$) to be $G\times M_{v}$. While a
large G would aid us in better handling of skew and computation as the
\granules{} can be processed in parallel, it comes at the cost of
significant scheduling and storage overheads. We empirically study the
sensitivity to $G$ (in \secref{subsec:sa}); based on this, our prototype uses $G = 24$. 

\label{ss:runtime_changes}

\noindent 
{\bf New locations for existing \granules{}:} 
Data generation patterns can significantly vary across different
stages, and jobs, due to heterogeneous
compute logics and data skew.
Thus a job $j$ may run out of its quota $Q_{j}$ on machine $m$,
leaving no room to grow already-materialized \granules{} of $j$ on
$m$. Thus, DS-M periodically reacts to runtime changes by determining,
$\forall j$: (1) which machines are at risk of being overloaded; (2)
which \granules{} on these machines to spread at other locations; and
(3) on which machines to them spread to.

Given a job $j$, DS-M invokes \circled{h5} to determine machines where
$j$ is using $\ge 75\%$ of its quota $Q_{j}$.
DS-M then starts {\em closing} some \granules{} of $j$ on these
machines; future intermediate data for these is materialized on
another machine, thereby mitigating potential
hotspots. Specifically, DS-M invokes \circled{h6} to pick \granules{}
that are either significantly larger in size or have a higher size
increase rate than others for $j$ on $m$. 
These \granules{} are more likely to dominate the load and potentially
violate $Q_{j}$. Focusing on them bounds the number of \granules{}
that will spread out.
DS-M groups the \granules{} selected based on the stage which
generated them, and invokes heuristic \circled{h3} as before to
compute the set of machines where to spread. Grouping helps to
maximize data locality,
and \circled{h3} provides load balance and fault tolerance.
\section{Data Visibility}
\label{sec:data_vis}

We now describe how \name{} offers run-time programmable data
monitoring (\secref{s:monitor}), and data-driven computation using
events (\secref{s:da}).

\subsection{Data Monitoring}
\label{s:monitor}

\name{} consolidates a \granule{} at one or a few locations
(\secref{s:data_format}). Thus, \granules{} can be analyzed in
isolation, simplifying data visibility.  We achieve scalable
monitoring via per-job masters (DS-JMs) which track light-weight stats
at the \granule{}-level.

\name{} supports both {\em built-in} and {\em customizable} modules that
periodically gather statistics per \granule{}, spanning properties of keys and
values. These statistics are carried to the DS-JM where they are aggregated
before being used by ES to take further data-driven actions (\secref{s:da}).

Built-in modules constantly collect light-weight statistics such as
current \granule{} size, number of (k,v) pairs and rate of growth; in
addition to supporting user programs, these are used by the store in
runtime data organization (\secref{ss:runtime_changes}).  Custom
modules are UDFs (user defined functions). Since supporting arbitrary
UDFs can impose high overhead, we restrict UDFs to those that can
execute in linear time and O(1) state. We provide a library of common
UDFs, such as computing the number of entries for which values are $<,
=,$ or $>$ than a threshold.

\subsection{Acting on Monitored Data Properties} \label{s:da}

\name{} supports data-driven computation via two key
abstractions - {\em events} and {\em ready triggers}. The decoupled
data and execution services interact with each other via events which
trigger, and track progress of, data-driven computation. Ready
triggers enable the DS-JM to decide when  \granules{}
can be deemed ready for corresponding computation to be run on them.

\begin{figure}[!t]
	\centering
	\includegraphics[clip,width=0.9\columnwidth]{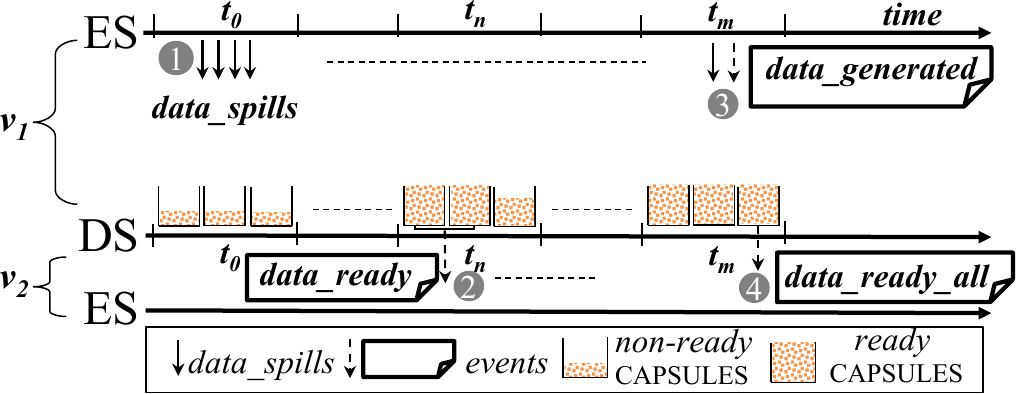}
	\caption{Data-driven computation facilitated by \name{} events. (1) Intermediate data ($v_{1}$) batches sent from ES to DS. (2) DS detects that 2 \granules{} are ready and sends data\_ready event from DS to ES leading to downstream computation ($v_{2}$). (3) ES sends data\_generated event to DS when entire output of $v_{1}$ pushed to DS. (4) DS sends data\_ready\_all event to ES indicating that all data\_ready events have been sent.}
	\label{fig:data_events}
		\vspace{-0.2in}
\end{figure}

\noindent
{\bf Events:}
\name{} introduces 3 types of events: (1) A $data\_ready$ event is
sent by the DS-JM to the ES whenever a \granule{} becomes ready (as
per the ready trigger definition) to trigger corresponding
computation. (2) A $data\_generated$ event is sent by the ES to the
DS-JM when a stage in the user program finishes generating all its
intermediate output. This event is required because in \name{} the
DS-JM is unaware of the number of tasks that the ES launches
corresponding to a stage, and thus cannot determine when a stage is
completed. (3) A $data\_ready\_all$ event is sent by the DS-JM to the
ES when a stage has generated all its intermediate data which is
ready to be consumed by an immediate downstream stage. 
This event is required as the ES is unaware of the status of
\granules{} (the total number of \granules{}, and whether they are done
being materialized). 


The use of these events is exemplified in
Fig.~\ref{fig:data_events}. Here: \circledd{1} when a stage $v_{1}$
generates a batch of intermediate data, a $data\_spill$ containing the
data is sent to the data store, which accumulates it into \granules{}
({\em $t_{0}$} through {\em $t_{m}$}). \circledd{2} Whenever the DS-JM
determines that a collection of $v_{1}$'s \granules{} (2 \granules{}
in Fig.~\ref{fig:data_events} at {\em $t_{n}$}) are ready for further
processing, it sends a $data\_ready$ event per \granule{} to the ES;
the ES could launch tasks of a consumer stage $v_{2}$ to process such
capsules.  This event carries per-\granule{} information such as: a
list of machine(s) on which each \granule{} is spread, and a list of
(aggregated) statistics (counters) collected by the {\em built-in}
data modules. \circledd{3} Finally, a $data\_generated$ event -- from
the ES, generated upon $v_1$ computation completing -- notifies the
DS-JM that $v_{1}$ finished generating $data\_spills$.  \circledd{4}
Subsequently, DS-JM notifies the ES via the $data\_ready\_all$ event,
that all \granules{} corresponding to $v_{1}$ have sent their
data\_ready events and have completely materialized (at {\em
  $t_{m}$}). This enables the ES to
determine when the immediate downstream stage $v_{2}$, that is reading
the data generated by $v_{1}$, has received all of its input data.


\noindent \textbf{Triggers:} The interaction between ES and DS via
events is enabled by triggers whose logic is based on
statistics collected by the data modules. A \name{} program can
provide {\em custom ready triggers} for each job stage, which the
\name{} client transfers to the DS-JM. \name{} also implements a {\em
  default ready trigger} which is otherwise applied.

 {\em Default ready trigger:} Here, the DS-JM deems
\granules{} ready when the computation generating them is done; this
is akin to a barrier in batch systems today and bulk synchronous
execution in graph analytics.  For a streaming job, the DS-JM deems a
\granule{} ready when it has $\ge X$ records ($X$ is a system
parameter that a user can configure) from producer tasks, and sends a
$data\_ready$ event to ES. On receiving this, ES executes a consumer
stage task on this \granule{}. This is akin to micro-batching in
existing streaming systems~\cite{spark-streaming}, with the crucial
difference that the micro-batch is not wall clock time-based, but is
based on the more natural intermediate data count.


{\em Custom ready trigger:} If available, the DS-JM deems
\granules{} as ready using custom triggers. Programmers define these
triggers based on knowledge about the semantics of the computation
performed, and the type of data properties sought.

Consider the partial execution of a batch (or graph) analytics job,
consisting of the first two logical stages (likewise, first two
iterations) $v_{1} \rightarrow v_{2}$. If the processing logic in
$v_{2}$ contains commutative+associative operations (e.g., sum, min,
count, etc.), it can start processing its input before all of it is in
place.  For this, the user can define a {\em pipelining ready
  trigger}, and instruct DS-JM to consider a \granule{} generated by
$v_{1}$ ready whenever the number of records in it reaches a threshold
$X$. This enables the ES to overlap $v_{2}$'s computation with
$v_{1}$'s data generation as follows: (1) Upon receiving a
$data\_ready$ event from the DS-JM for \granules{} which have $> X$
records, the ES launches tasks of $v_{2}$. (2) Tasks read the current
data, compute the associative+commutative function on the (k,v) data
read, and push the result back to data store (in the same \granules{}
advertised through the received $data\_ready$ event). (3) The DS-JM
waits for each \granule{} to grow back beyond threshold $X$ for
generating subsequent $data\_ready$ events. (4) Finally, when a
$data\_generated$ event is received from $v_{1}$, the DS-JM triggers a
final $data\_ready$ event for all the \granules{} generated by
$v_{1}$, and a subsequent $data\_ready\_all$ event, to enable
$v_{2}$’s final output to be written in \granules{} and fully consumed
by a downstream stage, say $v_{3}$ (similar to
Fig.~\ref{fig:data_events}). Such pipelining speeds up jobs in {\em
  batch} and {\em graph} analytics, as we show
in ~\secref{subsec:eval-graph-processing}. 

The above pipelining ready trigger can be extended {\em across}
\granules{}. E.g., the DS-JM could deem all \granules{} ready when the
number of entries generated across all \granules{} crosses a
threshold. In streaming, such triggers help improve efficiency and
performance (\secref{subsec:eval-stream-processing}). In a two-stage
streaming job $v_{1} \rightarrow v_{2}$, $v_{2}$ may (re)compute a
weighted moving average whenever $100$ data points with distinct keys
are generated by $v_{1}$. A pipelining trigger can be used to signal
all $v_{1}$ \granules{} as ready once said threshold is met.  This
brings {\em data driven-ness} to stream analytics: computation is
performed only when the required records have streamed into the
system.

In a similar manner, we can trigger processing based on {\em special
  records}. Stream processing systems often rely on ``low watermark''
records to ensure event-time processing~\cite{flink,streams}, and to
support temporal joins~\cite{streams}. Custom triggers can be used to
launch, {\em on demand}, temporal operators whenever a low watermark
record is observed at any of a stage's output \granules{}.
\section{Execution Service}
\label{sec:runtime_management}

The ES launches a per-job master (ES-JM) which late-binds the job’s
computation, i.e., given intermediate data ready for processing, and
available resources\footnote{Similar to existing frameworks, a
  cluster-wise Resource Manager decides available resources as per
  cross-job fairness}, (a) it determines optimal parallelism and
deploys tasks to minimize skew and shuffle, and (b) it maps
\granules{} to tasks in a resource-aware fashion
(\secref{es:formulation}). The ES-JM design naturally mitigates
stragglers (\secref{subsec:straggler_mitigation}), and facilitates
data-driven compute logic changes (\secref{subsec:runtime_changes}).

\subsection{Task Parallelism, Placement, and Sizing}
\label{es:formulation}

Given a set of ready \granules{} ($C$) for a stage, the ES-JM maps subsets of
\granules{} to tasks, and determines the location (across machines $M$) and the
size of the corresponding tasks based on available resources. This
multi-decision  problem, which evens out data volume processed by tasks in a
stage, and minimizes shuffle subject to resource contraints on each machine, can
be cast as a binary ILP (omitted for brevity).
However, the formulation  is non-linear;
even a linear version is slow to solve at scale.
For tractability, we propose an {\em iterative procedure} that applies
a set of heuristics (Tab.~\ref{tab:es_heuristics}) repeatedly until
tasks for all ready \granules{} are allocated, and their locations and
sizes (resources) determined.

\begin{table}[!t]
	\centering
	\begin{scriptsize}
		\begin{tabular}{c|c}
			\circled{h7} & \multicolumn{1}{|l}{
				\begin{tabular}[l]{@{}l@{}}
					\textit{// $\myvec{C}$: subsets of unprocessed \granules{}.}
					\\ a. $CaMax$ = $2 \times |c|$, $c$ is largest \granule{} $\in$ $C$;
					\\ b. Group all \granules{} $\in$ $C$ into subsets in strict order:
					\\ \hspace{0.22cm} i. data local \granules{} together;
					\\ \hspace{0.16cm} ii. each spread \granule, along data-local \granules{} together;
					\\ \hspace{0.10cm} iii. any remaining \granules{} together;
					\\ \hspace{0.02cm} subject to:
					\\ \hspace{0.16cm} iv. each subset size $\leq CaMax$;
					\\ \hspace{0.22cm} v. conflicting \granules{} don't group together;
					\\ \hspace{0.16cm} vi. troublesome \granules{} always group together. \\
			\end{tabular}} \\
			\cline{1-2}
      \circled{h8}& \multicolumn{1}{|l}{
				\begin{tabular}[l]{@{}l@{}}
					\textit{// $\myvec{M}$: preferred machines to process each subset $\in$ $\myvec{C}$.}
					\\ c. no machine preference for troublesome subsets $\in$ $\myvec{C}$
					\\ d. for every other subset $\in$ $\myvec{C}$ pick machine $m$ such that:
					\\ \hspace{0.1cm} i. all \granules{} in the subset are only materialized at $m$;
					\\ \hspace{0.05cm} ii. otherwise $m$ contains the largest materialization of the subset.
			\end{tabular}} \\
			\cline{1-2}
			\circled{h9} & \multicolumn{1}{|l}{
				\begin{tabular}[l]{@{}l@{}}
					Compute $\myvec{R}$: resources needed to execute each subset $\in$ $\myvec{C}$:
					\\ \hspace{0.1cm} e. $\myvec{A}$ = available resources for $j$ on machines $\myvec{M}$;
					\\ \hspace{0.1cm} f. $F$ = $\min{(\frac{\myvec{A}[m]}{ \text{total size of \granules{} allocated to $m$} }}$, for all $m \in \myvec{M}{)}$;
					\\ \hspace{0.1cm} g. for each subset $i \in \myvec{C}$:
					\\ \hspace{0.4cm} $\myvec{R}[i]$ = $F \times \text{total size of \granules{} allocated to $\myvec{C}[i]$}$.
			\end{tabular}} \\
		\end{tabular}
		\caption{Heuristics to group \granules{} and assign them to tasks}
		\label{tab:es_heuristics}
                \vspace{-0.2in}
	\end{scriptsize}
\end{table}

The iterative procedure consists of 3 steps - (a) generate optimal
subsets of \granules{} to minimize cross-subset skew (using
\circled{h7}), (b) decide on which machine should a subset be
processed to minimize shuffle (\circled{h8}), and (c) determine
resources required to process each subset (\circled{h9}).

First, we group \granules{}, $C$, into a collection of {\em subsets},
$\myvec{C}$, using \circled{h7}. We then try to assign each group to a
task. Our grouping into subsets attempts to ensure that data in a
subset is spread on just one or a few machines (lines (b.i-b.iii)),
which minimizes shuffle, and that the total data is spread roughly
evenly across subsets (line (b.iv)) making cross-task performance
uniform. We place a bound $CaMax$, equaling twice the size of the
largest \granule{}, on the total size of a subset (see line
(a)). This bound ensures that multiple (atleast 2)
\granules{} are present in each subset, which helps in mitigating
stragglers (\secref{subsec:straggler_mitigation}).

Second, we determine a preferred machine to process each subset using
\circled{h8}; this is a machine where most if not all \granules{} in the
subset are materialized (line (d)). Choosing a machine in this manner minimizes shuffle. 

Finally, given available resources across the preferred machines (from
the cluster-wide resource manager~\cite{yarn}) we need to allocate
tasks to process subsets. But some machines may not have resource
availability. For the rest of this iteration, we ignore such machines
and the subsets of \granules{} that prefer such machines.

Given machines with resources $\myvec{A}$, we assign a task for each
subset of \granules{} which can be processed, and allocate task
resources \emph{altruistically} using \circled{h9}.  That is, we first
compute the minimum resource available to process unit data ($F$; line
(f)). Then, for each task, the resource allocated (line (g)) is $F$
times the total data in the subset of \granules{} 
allocated to the task ($|\myvec{C}[i]|$). 

Allocating resources proportional to input size coupled with roughly
equal subset sizes, ensures that tasks have roughly equal finish times
in processing their subsets of \granules{}. Furthermore, by allocating
resources corresponding to the minimum available, our approach
realizes {\em altruism}: if a job gets more resources than what is
available for the most constrained subset, then it does not help the
job's completion time (because completion time depends on how fast the
most constrained subset is processed).  Altruistically ``giving back''
such resources helps speed up other jobs or other stages in the same
job. 

The above 3 steps repeat whenever new \granules{} are ready, or
existing ones can't be scheduled.
Similar to delay
scheduling~\cite{delay-scheduling}, we attempt several tries to
execute a group which couldn't be scheduled on its preferred machine
due to resource unavailability, before marking \granules{} {\em
  conflicting}. These are re-grouped in the next iteration (line
(b.v)). Finally, \granules{} that cannot be executed under any
grouping are marked troublesome (line (b.vi)) and processed
on any machine (line (d.ii)).

\subsection{Handling Stragglers}
\label{subsec:straggler_mitigation}

Data organization into \granules{} and late-binding computation to
data enables a natural, simple, and effective straggler mitigation
technique.  If a task struck by resource contention makes slower
progress than others in a stage, given that there are multiple
\granules{} assigned to a task, the ES-JM simply splits the task's
\granule{} group into two, in proportion of the task's processing
speed relative to average speed of other tasks in the stage.  It then
assigns the larger-group \granules{} to a new task, and places the
task using the approach above. This addresses stragglers via {\em work
  reallocation}, as opposed to using clone
tasks~\cite{late,mantri,rope,skewtune,dolly} in compute-centric
frameworks. Cloning waste resources and duplicates work.

\subsection{Runtime DAG Changes}  \label{subsec:runtime_changes}

Visibility into data enables run-time changes to tasks' logic. In
\name{}, we introduce {\em statistics} and {\em status} events,
defined next, to support this. Whenever a \granule{} is deemed ready,
the DS-JM sends a {\em statistics} event to the \name{} client, which
carries the \granule{} statistics. 
{\em Status} events help the
ES-JM query the \name{} client to check if the user program requires
alternate logic to be launched based on observed statistics.

Upon receiving \textit{data\_ready} events from the DS-JM, ES-JM sends
\textit{status} events to the \name{} client to determine
how to process a given \granule{}. Given
the user-provided compute logic and \granule{} statistics obtained
through \textit{statistics} events, the \name{} client notifies
the ES to take one the following actions: (1) \textit{no new action}
-- assign computation as planned; (2) \textit{ignore} -- don't perform
any computation, as the user program deemed the \granule{} to not have
any useful data to compute on;
(3) \textit{replace computation} with new logic
supplied by the user.
\section{Fault Tolerance}
\label{sec:ft}

\noindent {\bf Task Failure.} When a \name{} task fails due to a
machine failure, only the failed tasks need to be re-executed if the
input \granules{} are not lost. However, this will result in duplicate
data in all \granules{} for the stage leading to intermediate data
inconsistencies. To address this, we use checksums at the consumer
task-side \name{} library to suppress duplicate spills.

However, if the failed machine also contains the input \granules{} of
the failed task, then the ES-JM triggers the execution of the upstream
stage(s) to regenerate the input \granules{} of the failed
task. Recall that \name{}'s fault tolerance-aware \granule{} storage
(\secref{sec:data_org}) helps control the number of upstream
(ancestor) stages that need to be re-executed in case of data loss.

\noindent {\bf DS-M/DS-JM/ES-JM.}  \name{} maintains replicas of
DS-M/DS-JM daemons using Apache Zookeeper~\cite{zookeeper}, and fails
over to a standby. Given that \name{} decides the task composition of a
job at runtime in a data-driven manner, upon ES-JM failure, we simply
need to restart it so that it can resume handling events from the
DS. During this time already launched tasks continue to run.

\section{Implementation}
\label{sec:implementation}

We prototyped \name{} by modifying Tez~\cite{tez} and leveraging
YARN~\cite{yarn}. \name{}'s core components are application agnostic
and support diverse analytics as shown in \secref{sec:evaluation}.

The DS was implemented from scratch and consists of three kinds of
daemons: cluster-wide master (DS-M) and per-job masters (DS-JM), which
we discussed in earlier sections, and workers (DS-W). The DS-W runs on cluster machines and conducts node-level data
management. It handles storing the data received from the ES or from
other DS-Ws in a local in-memory file system
(\texttt{tmpfs}~\cite{tmpfs}) and transfers data to other DS-Ws per
DS-M directives. It collects statistics and reports to the DS-M/DS-JMs
via heartbeats. Finally, it provides ACKs to ES tasks for the data
they write. We use YARN to launch/terminate daemons.

The ES was implemented by modifying components of Tez to enable
data-driven execution. ES tasks are modified Tez tasks that have an interface to the
local DS-W as opposed to local disk or cluster-wide storage. The
\name{} client is a standalone process per-job. 

All communication (asynchronous) between DS, ES and \name{} client is
implemented through RPCs in YARN using Google
Protobuf~\cite{protobuf}. We also use RPCs for communication between
the YARN Resource Manager (RM) and ES-JM to propagate job resource
allocation (\secref{sec:runtime_management}).

\section{Evaluation}
\label{sec:evaluation}

We evaluated \name{} on a $50$-machine cluster deployed on
CloudLab~\cite{cloudlab} using publicly available benchmarks -- {\em
  batch} TPC-DS jobs, PageRank for {\em graph analytics}, and
synthetic {\em streaming} jobs. Unless otherwise specified, we set
\name{} to use default ready triggers, equal storage quota ($Q_{j}
=2.5GB$) and 24 \granules{} per machine.

\begin{figure}[!t]
	\centering
	\subfloat[][]{%
		\label{fig:running_jobs_batch}%
		\includegraphics[scale=0.3]{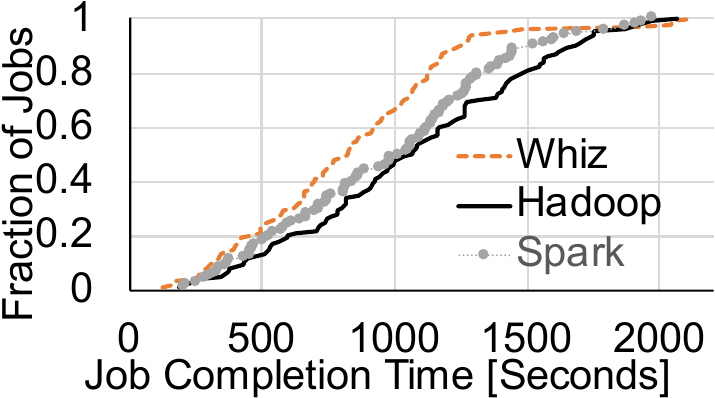}
	}
	\hspace{0.2cm}
	\subfloat[][]{%
		\label{fig:cdf_jobs_batch}%
		\includegraphics[scale=0.3]{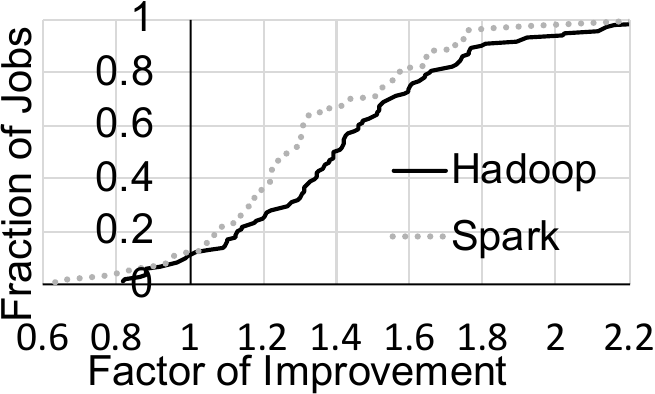}
	}
	\vspace{-2ex}
		\subfloat[][]{%
			\label{fig:running_tasks_batch}%
			\includegraphics[scale=0.3]{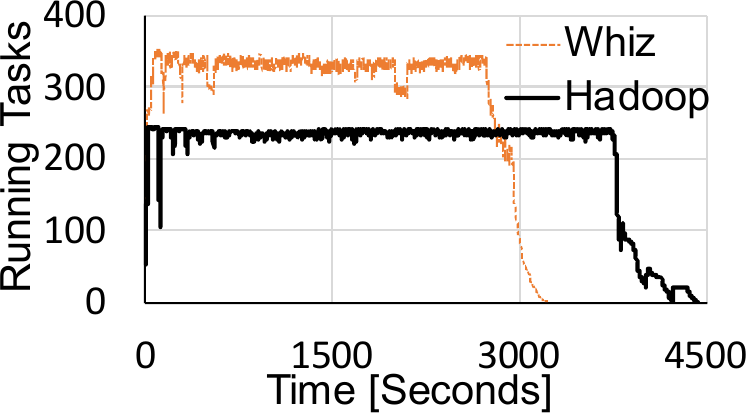}
		}
		\hspace{0.2cm}
		\subfloat[][]{%
			\label{fig:machine_cluster_load_batch}%
			\includegraphics[scale=0.29]{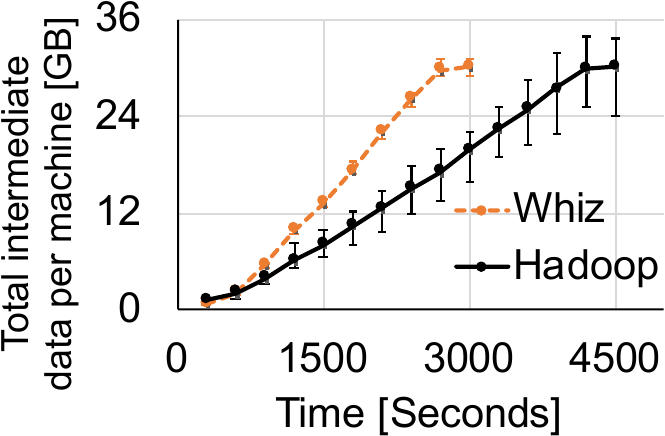}
		}
		\vspace{-2ex}
	\caption{(a) CDF of JCT; (b) CDF of factors of improvement of individual jobs using \name{} \wrt{} baselines~\cite{hadoop, spark}; (c) Running tasks; (d) Per machine \emph{average}, \emph{min} and \emph{max} storage load.}
	\label{fig:cluster_jobs_batch}
	\vspace{-0.6cm}
\end{figure}

\subsection{Testbed Experiments, Different Applications}
\label{subsec:ob}
\label{subsec:methodology}

\noindent
{\bf Workloads:} We consider a mix of jobs, all from TPC-DS ({\bf
  batch}), or all from PageRank ({\bf graph}). In each experiment, jobs
are randomly chosen and follow a Poisson arrival distribution with
average inter-arrival time of 20s. Each job lasts up to 10s of
minutes, and takes as input tens of GBs of data.
For {\bf streaming},
we created a synthetic workload from a job which periodically replays
GBs of text data from HDFS and returns top 5 most common words for the
first 100 distinct words found. We run each experiment thrice and 
present the median.

\noindent {\bf Cluster, baseline, metrics:} Our machines have 8 cores,
64GB of memory, 256GB storage, and a 10Gbps NIC. The cluster network
has a congestion-free core.  We compare \name{} as follows: (1) {\bf
  Batch}: {\em vs.} Tez~\cite{tez} running atop YARN~\cite{yarn}, for
which we use the shorthand ``Hadoop'' or ``CC''; and {\em vs.}
SparkSQL~\cite{sparksql}; (2) {\bf Graph}: {\em vs.}
Giraph (i.e., open source Pregel~\cite{pregel}); and {\em vs.}
GraphX~\cite{graphx}; (3) {\bf Streaming}: {\em vs.}
SparkStreaming~\cite {spark-streaming}. We study the relative
improvement in the average job completion time (JCT), or
Duration$_{CC}$/Duration$_{\name}$. We measure efficiency using
makespan. For a fair comparison with Hadoop and Giraph, we ensure that
they use \texttt{tmpfs}.


\subsubsection{Batch Analytics}

\noindent
{\bf Performance and efficiency:} Fig.~\ref{fig:running_jobs_batch}
shows the JCT distributions of \name{}, Hadoop, and Spark for the
TPC-DS workload. Only $0.4$ ($1.2$) highest percentile jobs are worse
off by $\le 1.06\times$ ($\le 1.03\times$) than Hadoop (Spark). \name{} speeds
up jobs by $1.4\times$ ($1.27\times$) on average, and $2.02\times$ ($1.75\times$) at
$95th$ percentile \wrt{} Hadoop (Spark). Also, \name{} improves
makespan by $1.32\times$ ($1.2\times$).

Fig.~\ref{fig:cdf_jobs_batch} presents improvement for individual
jobs. For more than $88\%$ jobs, \name{} outperforms 
Hadoop and Spark. Only $12\%$ jobs slow down to
$\le 0.81\times$ ($0.63\times$) using \name{}. 
Gains are $>1.5\times$ for $>35\%$ jobs.

\noindent {\bf Sources of improvements:} We observe that {\em more rapid
  processing}, and {\em better data management} contribute most to benefits.

First, we snapshot the number of running tasks across all the jobs in
one of our experiments when running \name{} and Hadoop
(Fig.~\ref{fig:running_tasks_batch}).
\name{} has $1.45\times$ more tasks scheduled over time which
translates to jobs finishing $1.37\times$ faster. It has $1.38\times$ better cluster
efficiency than Hadoop. Similar observations hold for Spark (omitted).

The main reasons for rapid processing/high efficiency are: (1) The DS
(\secref{s:solution}) ensures that most tasks are data local ($76\%$
in our expts).
This improves average {\em consumer} task completion time by
$1.59\times$. 
Resources thus freed can be used by other jobs' tasks. (2) Our ES can
provide similar input sizes for tasks in a stage
(\secref{es:formulation}) -- within $14.4\%$ of the mean (more in
\secref{subsec:be}).

Second, Fig.~\ref{fig:machine_cluster_load_batch} shows the size of
the cross-job total intermediate data  per
machine. We see that Hadoop generates heavily
imbalanced load spread across machines. This creates many storage
hotspots and slows down tasks competing on those machines. Spark is
similar.  \name{} mitigates hotspots (\secref{sec:data_org}) improving
overall performance.

\noindent {\bf \name{} slowdown:} We observe jobs generating less
intermediate data are more prone to performance losses, especially
under ample resource availability. A reason is that \name{} strives
for \granule{}-local task execution (\secref{es:formulation}).
If resources are unavailable, \name{} will assign the task to
a data-remote node, or get penalized waiting for data-local
placement. 
Also, \name{} gains are lower \wrt{} Spark. This is an
artifact of our Hadoop-based implementation, and of using a non-optimized
in-memory store.

Only $18\%$ of 
\granules{} across all jobs are spread across machines. Also, $> 25\%$ of the
jobs whose performance improves processed ``spread-out''
\granules{}; and $\le 14\%$ of the slowed jobs 
processed spread \granules{}.


\begin{figure}[!t]
	\centering
	\subfloat[][]{%
		\label{fig:running_jobs_graph}%
		\includegraphics[scale=0.27]{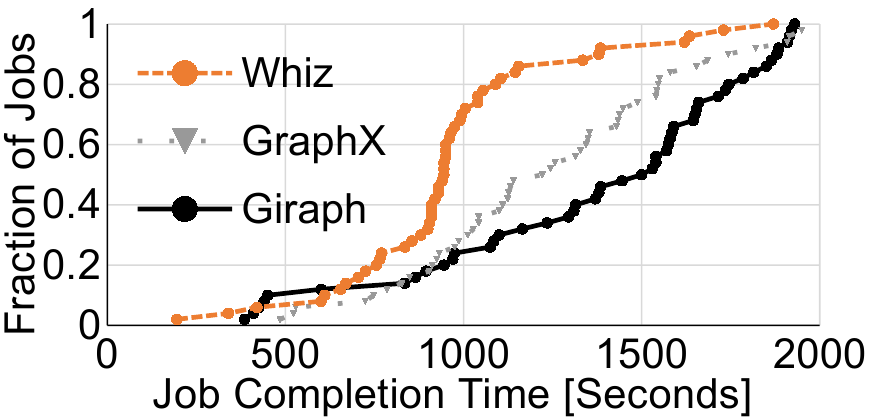}%
	}
	\hspace{0.2cm}
	\subfloat[][]{%
		\label{fig:cdf_jobs_graph}%
		\includegraphics[scale=0.27]{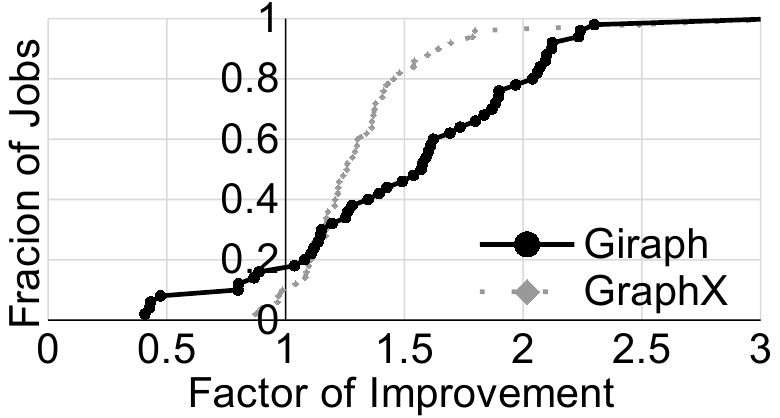}%
	}
	\vspace{-2ex}
	\subfloat[][]{%
		\label{fig:running_jobs_streaming}%
		\includegraphics[scale=0.28]{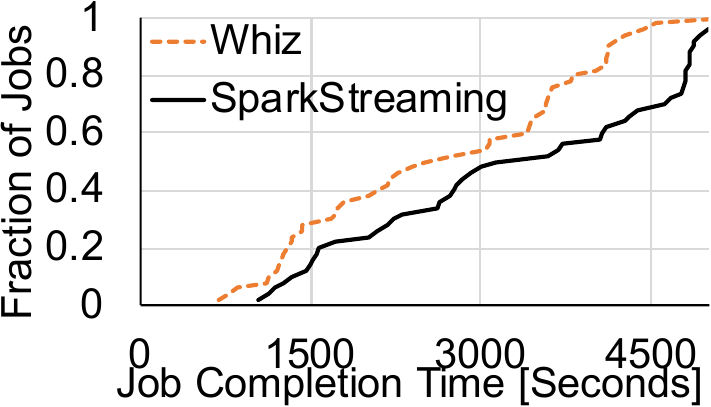}
	}
	\hspace{0.6cm}
	\subfloat[][]{%
		\label{fig:cdf_jobs_streaming}%
		\includegraphics[scale=0.28]{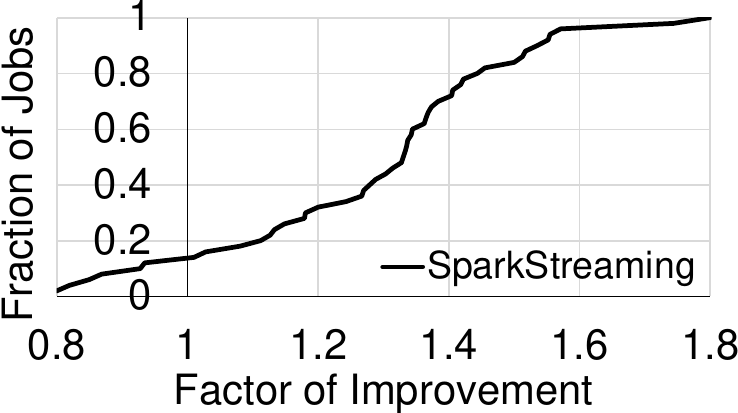}%
	}
	\vspace{-2ex}
	\caption{(a) CDF of JCT using \name{}, GraphX~\cite{graphx} and Giraph~\cite{giraph}; (b) CDF of factors of improvement of individual jobs using \name{} \wrt{} GraphX and Giraph; (c) CDF of JCT using \name{} and SparkStreaming~\cite{spark-streaming}; (d) CDF of factors of improvement of individual jobs using \name{} \wrt{} SparkStreaming.}
	\label{fig:cluster_jobs}
	\vspace{-0.6cm}
\end{figure}

\subsubsection{Graph Processing}
\label{subsec:eval-graph-processing}

We run multiple PageRank (40 iters.) jobs on the Twitter
Graph~\cite{tgr1, tgr2}. \name{} groups data (messages exchanged over
algorithm iterations) into \granules{} based on vertex ID. We use a
{\em custom ready trigger} (\secref{s:da}) so that a \granule{} is
processed only when $\ge 1000$ entries are present.

Fig.~\ref{fig:running_jobs_graph} shows the JCT distribution of
\name{}, GraphX and Giraph. \name{} speeds up jobs by $1.33\times$ ($1.57\times$) on average and $1.57\times$ ($2.24\times$) at the  $95th$ percentile {\wrt} GraphX (Giraph) (Fig.~\ref{fig:cdf_jobs_graph}). Gains are lower {\wrt} GraphX, due to its efficient implementation atop Spark. However, $< 10\%$ jobs are slowed down by $\le 1.13\times$.


Improvements arise for two reasons. First, \name{} is able to {\em
  deploy appropriate number of tasks only when needed}: custom ready
triggers immediately indicate data availability, and runtime
parallelism (\secref{es:formulation}) allows messages to high-degree
vertices~\cite{graphx} to be processed by more than one task.  Also,
\name{} has $1.53\times$ more tasks (each runs multiple
vertex programs) scheduled over time; rapid processing and runtime
adaptation to data directly leads to jobs finishing faster.
Second, because of triggered compute, \name{} doesn't
hold resources for a task if not needed, resulting in $1.25\times$ better
cluster efficiency.


\subsubsection{Stream Processing}
\label{subsec:eval-stream-processing}

We configure the Spark Streaming micro-batch interval to 1 minute. 
With \name{}, we implemented a {\em custom ready trigger} to enable
computation whenever $\ge 100$ distinct entries are present in the
intermediate data.  Figures~\ref{fig:running_jobs_streaming},
~\ref{fig:cdf_jobs_streaming} show our results. \name{} speeds up
jobs by $1.33\times$ on average and $1.55\times$ at the $95th$ \%-ile. Also,
$15\%$ of the jobs are slowed down to around $0.8\times$.

The reason for gains is {\em data-driven computation} via custom
  ready triggers; \name{} does not have to delay execution till the
next micro-batch if data can be processed now. A SparkStreaming task
has to wait as it has no data visibility. In our experiments, more
than $73\%$ executions happen at less than $40s$ time intervals with
\name{}.

\name{} suffers due to implementation atop a non-optimized
streaming stack. Launching tasks in YARN is significantly slower than
acquiring tasks in Spark. This can be exacerbated by ES stickiness to
data locality. However, this is amortized over long job run times.

\subsubsection{\name{} Overheads}
\noindent \textbf{CPU, memory overhead:} We find that DS-W
(\secref{sec:implementation}) processes inflate the memory and CPU
usage by a negligible amount even when managing data close to
storage capacity.
DS-M and DS-JM have similar resource profiles.

\noindent \textbf{Latency:}
We compute the average time to process heartbeats from various ES/DS daemons, and \name{}
client. For $5000$ heartbeats, the time to process each is
$2-5ms$. We implemented the \name{} client and ES-JM logic atop Tez AM. Our
changes inflate AM decision logic by $\le 14ms$ per request with 
negligible increase in AM memory/CPU.


\noindent
\textbf{Network overhead} from events/heartbeats is negligible.



\subsection{Benefits of Data-driven Computation}
\label{subsec:be}

The overall benefits of \name{} above also included the effects of
dynamic parallelism/placement/sizing (\secref{es:formulation}) and
straggler mitigation (\secref{subsec:straggler_mitigation}). 
We now delve deeper into them to shed more light on data-driven computation benefits during one of our TPC-DS runs.  

\begin{figure}[!t]
	\centering
	\subfloat[][]{%
		\label{fig:j1_skew}%
		\includegraphics[scale=0.25]{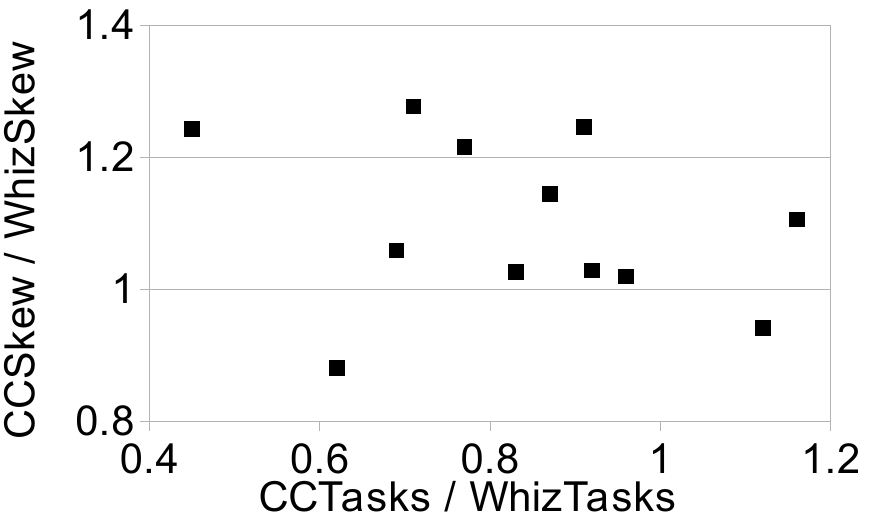}%
	}
	\hspace{0.2cm}
	\subfloat[][]{%
		\label{fig:j2_skew}%
		\includegraphics[scale=0.25]{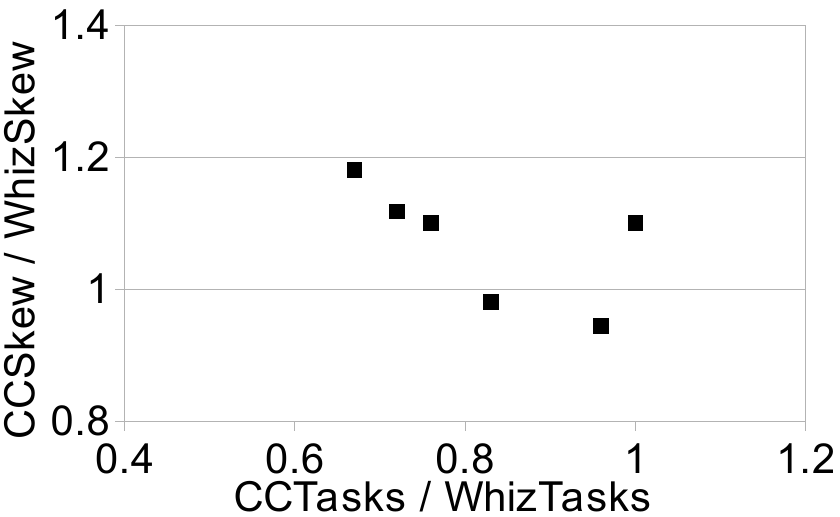}%
	}
	\vspace{-2ex}
	\caption{Fraction of tasks allocated by CC {\wrt} to \name{} {\vs} the fraction of skew among tasks due to CC {\wrt} \name{}, in a given stage: (a) for a job with $12$ stages where \name{} improves JCT by $1.6\times$; (b) for a job with $6$ stages, where \name{} improves JCT by $1.2\times$. $\frac{CCSkew}{WhizSkew} > 1$ means \name{} has less skew; $\frac{CCTasks}{WhizTasks} < 1$ means CC under-parallelize a given stage.}
	\label{fig:skew}
	\vspace{-0.6cm}
\end{figure}

\noindent
{\bf Skew and parallelism:} Fig.~\ref{fig:skew} shows fractions of skew and parallelism as generated by CC {\wrt} \name{} for two TPC-DS jobs from one of our runs. \name{}'s ability to dynamically change parallelism at runtime, driven by the number of \granules{} for each vertex, leads to significantly less data skew than CC. When CC is under-parallelizing, the skew is significantly higher than \name{} (up to $1.43\times$). Over-parallelizing does not help either; CC incurs up to $1.15\times$ larger skew, due to its rigid data partitioning and tasks allocation schemes. Even when \name{} incurs more skew (up to $1.26\times$), corresponding tasks will get allocated more resources to alleviate this overhead (\secref{es:formulation}).

\noindent
{\bf Straggler mitigation:} We evaluated the benefits enabled by \name{}'s straggler mitigation strategy {\wrt} (1) the default speculative execution strategy from CC and (2) when there is no strategy. We slowdown the same set of $10\%$ of machines in the cluster for each run. Table~\ref{table:stragglers_jct} shows our results. \name{}'s straggler mitigation strategy improves JCT $1.47\times$ on average {\wrt} no strategy. This is because stragglers can significantly impact overall performance. However, even when the CC speculative execution strategy is enabled, \name{} performs $1.19\times$ better. This is due to its ability to reassign only the unprocessed input data to the speculative tasks instead of launching clones of the straggler tasks. 


Additionally, we run microbenchmarks to delve further into \name{}'s data-driven benefits (results in Appendix~\ref{sec:appendix-microbenchmarks}).

\begin{figure}[!t]
	\centering
	\subfloat[][]{%
		\label{fig:dl_ft}%
		\includegraphics[scale=0.17]{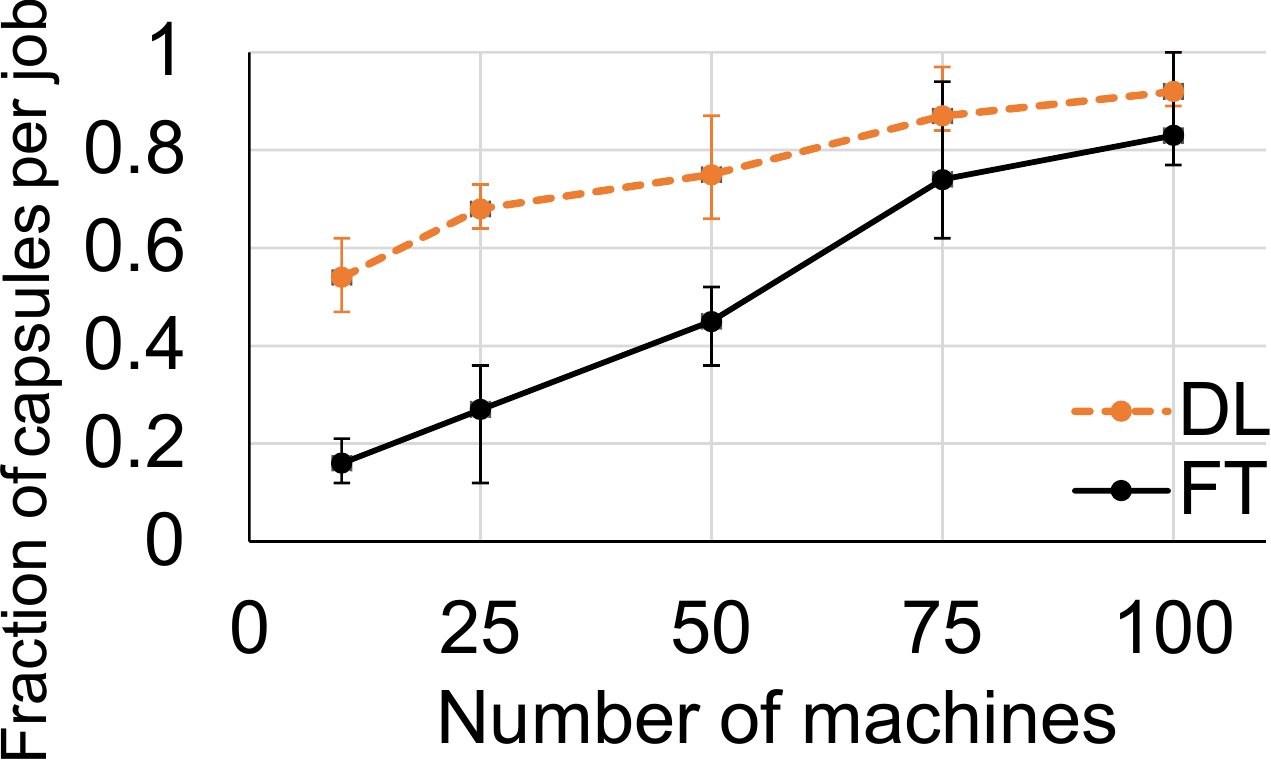}%
	}
	\hspace{0.2cm}
	\subfloat[][]{%
		\label{fig:lb}%
		\includegraphics[scale=0.3]{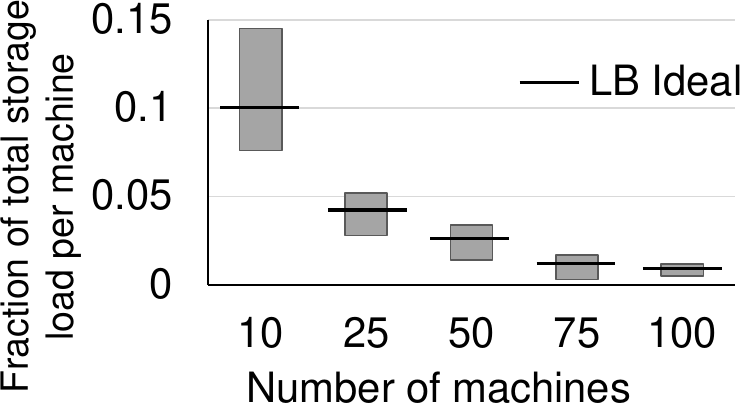}%
	}
	\vspace{-2ex}
	\caption{(a) Average, min and max fraction of \granules{} which are data local (DL) respectively fault tolerant (FT) across all the jobs for different cluster load; (b) Max, min and ideal storage load balance (LB) on every machine for different cluster load.}
	\label{fig:dl_ft_lb}
	\vspace{-0.6cm}
\end{figure}

\subsection{Load Balancing, Locality, Fault Tolerance}
\label{subsec:multiple_objectives}
To evaluate DS load balancing (LB), data locality (DL) and fault
tolerance (FT), we stressed the data organization under different
cluster load. We used job arrivals and all
stages' \granule{} sizes from one of our TPC-DS runs.


The main takeaways are (Fig.~\ref{fig:dl_ft_lb}): (1) \name{}
prioritizes DL and LB over FT across cluster loads
(\secref{s:solution}); (2) when the available resources are scarce
($5\times$ higher load than initial), all three metrics suffer. However,
the maximum load imbalance per machine is $<1.5\times$ than the ideal,
while for any job, $\ge 47\%$ of the \granules{} are DL. Also, on
average $16\%$ of the \granules{} per job are FT; (3) less cluster
load ($0.6\times$ lower than initial) enables more opportunities for
DS to maximize all of the objectives: $\ge 84\%$ of the per-job
\granules{} are DL, $71\%$ are FT, with at most $1.17\times$ load
imbalance per machine than the ideal.

\begin{table}
	\centering	
	 \setlength{\tabcolsep}{3pt}
	\subfloat[\vspace{0.8em}\label{table:failures_jct}%
	]{%
		\begin{minipage}{0.4\columnwidth}
			{\scriptsize
			\centering
			\begin{tabular}[b]{c|ccc}
				\multicolumn{1}{c|}{\% Machines} & \multicolumn{3}{c}{JCT [Seconds]}  \\
				\multicolumn{1}{c|}{Failed} & \textbf{Avg} & \textbf{Min} & \textbf{Max} \\
				\hline
				\textbf{None} & 725 & 215 & 2100 \\
				\hline
				\textbf{10\%} & 740 & 250 & 2320 \\
				\hline
				\textbf{25\%} & 820 & 310 & 2360 \\
				\hline
				\textbf{50\%} & 1025 & 350 & 2710 \\
				\hline
				\textbf{75\%} & 1600 & 410 & 3300 \\\hline
			\end{tabular}
			\vspace{-2em}			
		}
		\end{minipage}%
	}
\hspace{3em}	
 \subfloat[%
 \label{table:stragglers_jct}%
 ]{%
 	\begin{minipage}{0.4\columnwidth}
 		{\scriptsize
 		\centering
		\begin{tabular}[b]{c|ccc}
			\multicolumn{1}{c|}{Straggler} & \multicolumn{3}{c}{JCT [Seconds]}  \\
			\multicolumn{1}{c|}{Strategy} & \textbf{Avg} & \textbf{Min} & \textbf{Max} \\
			\hline
			\textbf{None} & 1200 & 410 & 2750 \\
			\hline
			\textbf{Speculative} & 970 & 280 & 2300 \\
			\hline
			\textbf{\name{}} & 815 & 330 & 1920 \\\hline
		\end{tabular}
 	}
 	\end{minipage}%
 }
\vspace{-1.5em}
		\caption{\name{}'s variation in JCT in the presence of: (a) random machines failures; (b) different straggler mitigation strategies; {\bf Speculative} is the default approach used by CC and {\bf \name{}} is our approach as described in ~\secref{subsec:straggler_mitigation}.}\label{whatever}
\vspace{-2.2em}
\end{table}

\noindent {\bf Failures:} Using the same workload, we also evaluated the performance impact in the presence of machine failures (Table~\ref{table:failures_jct}). We observe that \name{} does not degrade job performance by more than $1.13\times$ even when $50\%$ of the machines fail. This is mainly due to DS's ability to organize capsules to be FT across ancestor stages and avoid data recomputations. Even when $75\%$ of the machines fail, the maximum JCT does not degrade by more than $1.57\times$, mainly due to capsules belonging to some ancestor stages still being available, which leads to fast recomputation for corresponding downstream vertices.

\subsection{Sensitivity Analysis}
\label{subsec:sa}

\noindent
{\bf Impact of Contention:} We vary storage load, and hence resource
contention, by changing the number of machines while keeping the
workload constant; half as many servers lead to twice as much load. We
see that at $1\times$ cluster load, \name{} improves over CC by
$1.39\times$ ($1.32\times$) on average in terms of JCT
(makespan). Even at high contention (up to $4\times$), \name{}'s gains
keep increasing $1.83\times$ ($1.42\times$). This is because data is load balanced leading to {\em
  few storage hotspots}, {\em better isolation}, and \name{}
{\em minimizes resource wastage} and the time spent in shuffling.

\noindent
{\bf Impact of $G$ (number of \granules{} per machine):} We now provide the rationale for picking 
$G = 24$. Table~\ref{table:num_granules} shows the
factors of improvements \wrt{} CC for different values of $G$ and
levels of contention.

The main takeaways are as follows: for $G = 8$ the performance gap
between \name{} and CC is low ($< 1.1x$). This is expected because
small number of \granules{} results in less data locality (each
\granule{} is more likely to be spread). Further, the gap decreases
at high resource contention. In fact, at $4x$ the cluster load, CC
performs better ($0.85x$). At larger values of $G$ the performance gap
increases. For example, at $G = 24$, its gains are the most (between
$1.39x$ and $1.58x$). This is because larger $G$ implies (1) more
flexibility for \name{} to balance the load across machines; (2) more
likely that few \granules{} are spread out; (3) lesser data skew and
more predictable per task performance. However, a very large $G$ does
not necessarily improve performance, as it can lead to massive task
parallelism. The resulting scheduling overhead degrades performance, especially at high load.

\begin{table}
		\centering	
	{\scriptsize
	\vspace{-1em}
	\setlength{\tabcolsep}{3pt}
		\begin{tabular}[b]{c|cccccccc}
			\multicolumn{1}{c|}{Multiple of} & \multicolumn{8}{c}{ $\#$ Capsules}  \\
			Original Load & \textbf{8} & \textbf{16} & \textbf{20} & \textbf{24} & \textbf{28} & \textbf{32} & \textbf{36} & \textbf{40}\\
			\hline
			\textbf{1} & 1.07 & 1.33 & 1.46 & 1.52 & 1.57 & 1.63 & 1.54 & 1.46\\
			\hline
			\textbf{2} & 1.10 & 1.16 & 1.53 & 1.58 & 1.56 & 1.61 & 1.47 & 1.31\\
			\hline
			\textbf{4} & 0.85 & 1.12 & 1.34 & 1.39 & 1.32 & 1.16 & 0.95 & 0.74\\\hline
		\end{tabular}
			}
	\vspace{-0.5em}
	\caption{Factors of improvement \wrt{} CC for different number of \granules{} per machine and cluster load.} \label{table:num_granules}	
	\vspace{-2.2em}
\end{table}

\label{subsec:es_changes}

\noindent
{\bf Altruism:} Can altruistically deciding task sizing
(\secref{sec:runtime_management}) impact job performance? We compare
\name{}'s approach with a greedy task sizing approach, where each task
gets $\frac{1}{\#jobs}$ resource share per machine, and uses all of
it. For the same workload as ~\secref{subsec:multiple_objectives},
\name{} actually {\em speeds up} jobs by $1.48\times$ on avg., and
$3.12\times$ ($4.8\times$) at $75th$ \%-ile ($95th$ \%-ile)
\wrt{} greedy approach. Only $16\%$ jobs are slowed down by $\le
0.6\times$. Altriusm, by late-binding resources with the slowest potential task,
helps all jobs benefit.			
\section{Summary}
\label{sec:summary}

The compute-centric nature of existing frameworks hurts flexibility,
efficiency, isolation, and performance. \name{} reenvisions analytics
frameworks by cleanly separating computation from intermediate
data. Via programmable monitoring of data properties and a rich event
abstraction, \name{} enables data-driven decisions for what
computation to launch, where to launch it, and how many parallel
instances to use, while ensuring isolation. Our evaluation using
batch, stream and graph workloads shows that \name{} outperforms
state-of-the-art frameworks.

\phantomsection
\label{EndOfPaper}

{
	\raggedright
	\balance
	\bibliography{dcsplit}
	\bibliographystyle{abbrv}
}

\newpage

\appendix

\section{\name{} Sample Programs}
\label{sec:appendix-programs}

A key advantage of \name{} is its programming model wherein users no longer have to provide low level details and gain the benefits of data-driven computation using our APIs. While we showed a sample batch analytics application earlier (\secref{sec:programming-model}), we now show how to write more diverse application atop \name. More specifically, we show how to write a graph analytics job and a stream processing job. 

\begin{figure}
    \centering
    \includegraphics[clip,width=\columnwidth]{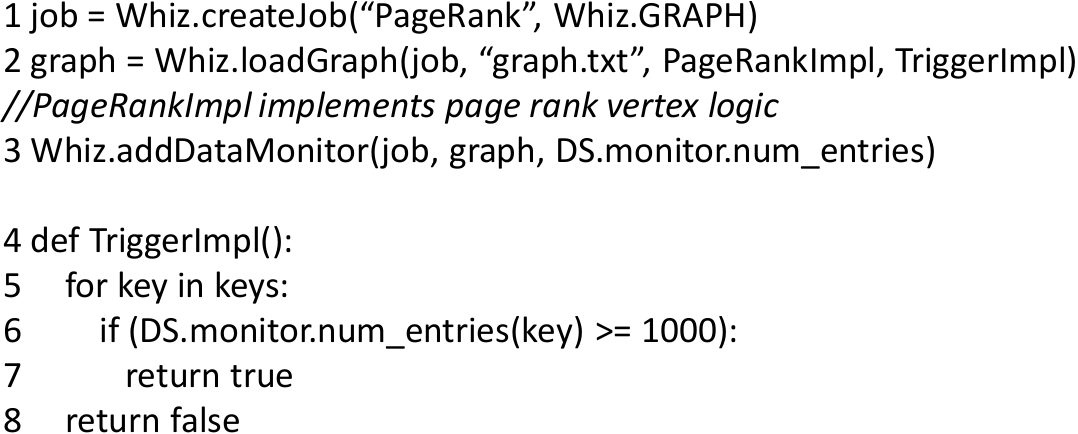}
    \vspace{-1em}
    \caption{An example graph job in \name{}}
    \label{fig:graph-program}
\end{figure}

\noindent
\textbf{Graph Analytics (Fig.~\ref{fig:graph-program}).} We consider the graph analytics application that we used in our testbed experiments (\secref{sec:evaluation}), .i.e., a job running the PageRank algorithm. Job composition details are similar to existing frameworks (lines 1-2). Similar to the $createStage$ API which is used for batch/stream jobs, we have a $loadGraph$ API which constructs the graph, given the input. Additionally, we specify the vertex program logic to run PageRank via the PageRankImpl function. 

Crucially, \name{} allows a user to run the graph algorithm in a data-driven manner by specifying the implementation of TriggerImpl. TriggerImpl specifies that as soon as 1000 entries corresponding to a vertex are seen, then we trigger the execution of the vertex program, leading to pipelined execution (lines 4-8). The user instructs the DS to use the built-in module to collect the per-key counts (line 3). This simple yet powerful trigger naturally deals with high-degree vertices and leads to better performance and efficiency (as seen in ~\secref{subsec:eval-graph-processing}).

\begin{figure}
    \centering
    \includegraphics[clip,width=\columnwidth]{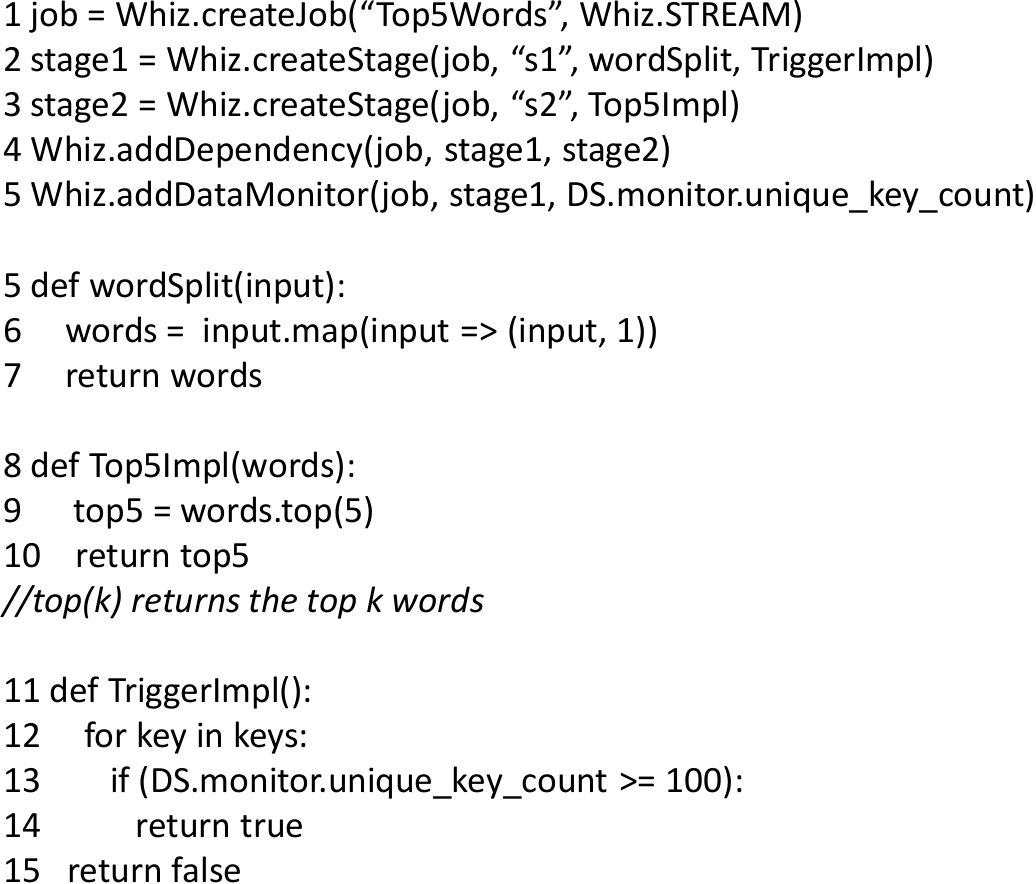}
    \vspace{-1em}
    \caption{An example streaming job in \name{}}
    \label{fig:stream-program}
\end{figure}

\noindent
\textbf{Stream Analytics. (Fig.~\ref{fig:stream-program})} We consider the stream analytics job that we used in our testbed experiments (\secref{sec:evaluation}), i.e., a job that return the top 5 words when we see 100 distinct words. As before, job composition details are similar to existing frameworks (lines 1-3). This job can be realized as a 2-stage job where in $S1$ processes the input words (lines 5-7) and $S2$ return the top 5 common words (lines 8-10). In this example, the user can leverage the flexibility of specifying a custom ready trigger to trigger computation. More specifically, the user can specify, using built-in data modules that trigger the $S2$ only when 100 distinct keys are encountered (lines 11-15). As we have seen in ~\secref{subsec:eval-stream-processing}, this trigger leads to better performance.

\section{Allocating \Granules{} to Machines ILP}
\label{s:formulation}

\begin{figure}
  \centering
  \begin{scriptsize}
    \begin{tabular}{r|p{6cm}}
      \multicolumn{2}{l}{\textbf{Objectives (to be minimized):}} \\
    $O_1$ & $\max_i  \left(\summation{k}{} (b^k_i + x^k_i e^k) \right)$ \\ \hline
    $O_2$ & $\summation{k}{} \left(  P^k - \left(\summation{i \in I_-^k}{}
    x_i^k\right) b_{\ihat(k)}^k
    + \summation{i \in I_+^k}{} x_i^k  \left(b_i^k + e^k \right) \right) $ \\ \hline
    $O_3$ & $\summation{k}{} \left( (1-f^k) \summation{i \in I^\circ}{} x_i^k
    \right)$ \\
      \multicolumn{2}{l}{\textbf{Constraints:}} \\
    $C_1$ & $\summation{k: J(g_k) = j}{} \left( b_i^k + x_i^k e^k \right) \leq
    Q_j , \quad \forall j, i$ \\
      \multicolumn{2}{l}{\textbf{Variables:}} \\
    $x_i^k$ & Binary indicator denoting capsule $g_k$ is placed on machine $i$
    \\
      \multicolumn{2}{l}{\textbf{Parameters:}} \\
    $b_i^k$ & Existing number of bytes of capsule $g_k$ in machine $i$ \\
      \hline
    $e^k$ & Expected number of remaining bytes for capsule, $g_k$ \\ \hline
    $P^k$ & $e^k + \summation{i}{} b_i^k$ \\ \hline
    $J(g_k)$ & The job ID for job $g_k$ \\ \hline
      $\ihat(k)$ &  $\textrm{argmax}_i \quad  b_i^k$ \\ \hline
      $I_-^k, I_+^k$ & $\{i: b_i^k  \leq b_{\ihat(k)}^k - e^k \}$, $\{i: b_i^k
      > b_{\ihat(k)}^k - e^k \}$\\ \hline
    $f^k$ & Binary parameter indicating that capsules  for same stage as $g_k$
      share locations with capsules for preceding stages \\ \hline
    $I^\circ$ & Set of machines where capsules of preceding stages are stored
      \\ \hline
    $Q_j$ & Administrative storage quota for job, $j$.
  \end{tabular}
  \end{scriptsize}
  \caption{Binary ILP formulation for \granule{} placement.
    \label{fig:formulation}}
  \vspace{-0.2in}
\end{figure}

We consider how to place multiple jobs' \granules{} to avoid
hotspots, reduce per-\granule{} spread (for data locality) and minimize
job runtime impact on data loss.  We formulate a
binary integer linear program (see Fig.~\ref{fig:formulation}) to this
end.
The indicator decision variables, $x_i^k$, denote that all future data
to \granule{} $g_k$ is materialized at machine $M_i$. The ILP finds
the best $x_i^k$'s that minimizes a multi-part weighted objective function, one
part each for the three objectives mentioned above.

The first part ($O_1$) represents the maximum amount of data stored
across all machines across all \granules{}. Minimizing this ensures
load balance and avoids hotspots.  The second part ($O_2$) represents
the sum of {\em data-spread penalty} across all \granules{}. Here, for
each \granule{}, we define the primary location as the machine with
the largest volume of data for that \granule{}. The total volume of
data in non-primary locations is the data-spread penalty, incurred
from shuffling the data prior to processing it.  The third part
($O_3$) is the sum of fault-tolerance penalties across \granules{}.
Say a machine $m$ storing intermediate for current stage $s$ fails;
then we have to re-execute $s$ to regenerate the data.  If the
machine also holds data for ancestor stages of $s$ then multiple
stages have to be re-executed. If we ensure that data from parent and
child stages are stored on different machines, then, upon child data
failure only the child stage has to be executed. We model this by
imposing a penalty whenever a \granule{} in the current stage is
materialized on the same machine as the parent stage. Penalties $O_2$,
$O_3$ need to be minimized.

Finally, we impose isolation constraint ($C_1$) requiring the total
data for a job to not exceed an administrator set quota $Q_j$. Quotas
help ensure isolation across jobs. 

However, solving this ILP at scale can take several tens of seconds delaying \granule{} placement. Thus, \name{} uses a linear-time rule-based heuristic to place capsules (as described in~\secref{sec:data_org}).





\section{\name{} Microbenchmarks}
\label{sec:appendix-microbenchmarks}

Apart from the experiments on the 50-machine cluster
(\secref{sec:evaluation}), we also ran several microbenchmarks to delve
deeper into \name{}’s data-driven benefits. The microbenchmarks were run on a
5 machine cluster and the workloads consists of the following jobs:
$\mathbb{J}_1$ ($v_{1} \rightarrow v_{2}$) and $\mathbb{J}_2$ ($v_{1}
\rightarrow v_{2} \rightarrow v_{3}$). These patterns typically occur in TPC-
DS queries.

\begin{figure}
    \centering
    \subfloat[][\vspace{-0.8em}]{%
        \label{fig:ctrl_parallelism}%
        \includegraphics[scale=0.3]{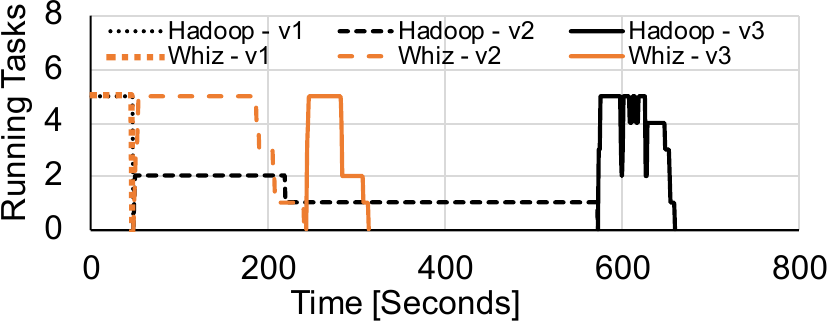}
    }
    \hspace{0.1cm}
    \subfloat[][\vspace{-0.8em}]{%
        \label{fig:straggler_mitigation}%
        \includegraphics[scale=0.3]{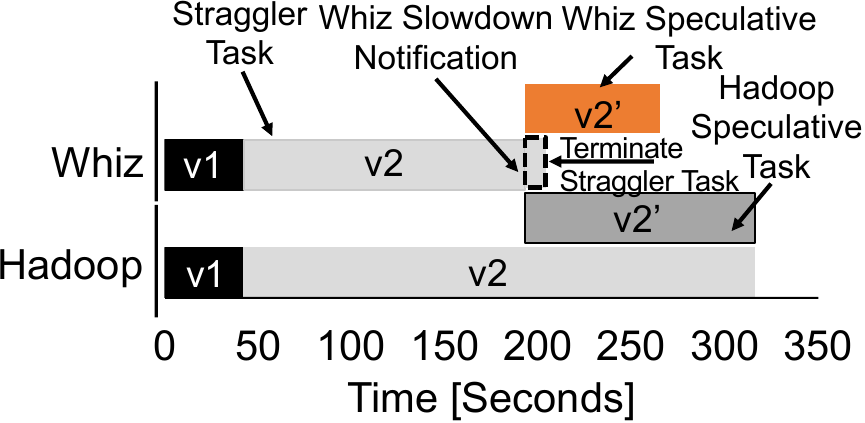}%
    }
    \vspace{-1ex}
    \caption{(a) Controlling task parallelism significantly
          improves \name{}'s performance over CC. (b) Straggler mitigation with \name{} and CC.}
    \label{fig:blah}
\end{figure}

\noindent
{\bf Skew and parallelism:} Fig.~\ref{fig:ctrl_parallelism} shows
the execution of one of the $\mathbb{J}_2$ queries from our workload when
running \name{} and CC. \name{} improves JCT by $2.67\times$ over CC. CC
decides stage parallelism tied to the number of data partitions. That
means stage $v_{1}$ generates $2$ intermediate partitions as
configured by the user and $2$ tasks of $v_{2}$ will process
them. However, execution of $v_{1}$ leads to data skew among the $2$
partitions ($1GB$ and $4GB$).
On the other hand, \name{} ends up generating \granules{} that are approximately equal in size and decides at
runtime a max. input size per task of $1GB$ (twice the largest \granule{}). This leads to running $5$
tasks of $v_{2}$ with equal input size and $2.1\times$ faster completion time of $v_{2}$ than CC.

Over-parallelizing execution does not help. With CC, $v_{2}$ generates
$12$ partitions processed by $12$ $v_{3}$ tasks. Under resource crunch,
tasks get scheduled in multiple waves (at $570$s in
Fig.~\ref{fig:ctrl_parallelism}) and completion time for
$v_{3}$ suffers ($85$s). In contrast, \name{} assigns at runtime only $5$
tasks of $v_{3}$ which can run in a
single wave; $v_{3}$ finishes $1.23\times$ faster.

%

\noindent
{\bf Straggler mitigation:} We run an instance of $\mathbb{J}_1$ with
$1$ task of $v_{1}$ and $1$ task of $v_{2}$ with an input size of
$1GB$. A slowdown happens at the $v_{2}$ task, which was assigned 2 \granules{} by \name. 


In CC (Fig.~\ref{fig:straggler_mitigation}), once a straggler is
detected ($v_{2}$ task at $203$s), it is allowed to continue, and a
speculative task $v'_{2}$ is launched that duplicates $v_{2}$'s
work. The work completes when $v_{2}$ {\em or} $v'_{2}$ finishes (at
$326$s).  In \name{}, upon straggler detection, the straggler ($v_{2}$)
is notified to finish processing the current \granule{}; a task
$v'_{2}$ is launched and assigned data from $v_{2}$'s unprocessed
\granule{}. $v_{2}$ finishes processing the first \granule{} at
$202$s; $v'_{2}$ processes the other \granule{} and finishes
$1.7\times$ faster than $v'_{2}$ in CC.

\noindent
{\bf Runtime logic changes:} We consider a job which processes words
and, for words with $<100$ occurrences, sorts them by frequency. The
program structure is $v_{1} \rightarrow v_{2} \rightarrow v_{3}$,
where $v_{1}$ processes input words, $v_{2}$ computes word
occurrences, and $v_{3}$ sorts the ones with $<100$ occurrences. In
CC, $v_{1}$ generates $17GB$ of data organized in $17$ partitions;
$v_{2}$ generates $8GB$ organized in $8$ partitions. Given this, $17$
$v_{2}$ tasks and $8$ $v_{3}$ tasks execute, leading to a CC JCT of
$220$s. Here, the entire data generated by $v_{2}$ has to be analyzed
by $v_{3}$. In contrast, \name{} registers a custom DS module to
monitor \#occurrences of all the words in the \granules{}
generated by $v_{2}$. We implement actions to \emph{ignore} $v_{2}$
\granules{} that don't satisfy the processing criteria of $v_{3}$
(\secref{subsec:runtime_changes}). At runtime, $2$ statistics events are
triggered by the DS module, and $6$ tasks of $v_{3}$ (instead of $8$)
are executed; JCT is $165$s ($1.4\times$ better).
\end{document}